\documentclass{sig-alternate}
\usepackage{times}
\usepackage{booktabs}
\usepackage[dvipsnames]{color}
\usepackage{subcaption}
\usepackage{xspace}
\usepackage{tikz}
\usepackage[export]{adjustbox}
\usepackage{enumitem}
\usepackage[numbers]{natbib}
\usepackage[hypertexnames=false,pdfpagelabels=false]{hyperref}

\newif{\ifshowcomments}
\showcommentstrue

\newcommand{\cut}[1]{}

\newcommand{\chenhao}[1]{}
\newcommand{\vn}[1]{}
\newcommand{\llee}[1]{}
\newcommand{\cd}[1]{}

\ifshowcomments
    \renewcommand{\chenhao}[1]{\textcolor{blue}{\small [CT: #1]}}
    \renewcommand{\vn}[1]{\textcolor{magenta}{\small [VN: #1]}}
    \renewcommand{\llee}[1]{\textcolor{pink}{\small [LL: #1]}}
    \renewcommand{\cd}[1]{\textcolor{Green}{\small [CD: #1]}}
\fi

\newcommand{\fullcmv}{{\tt /r/ChangeMyView}\xspace}
\newcommand{\plural}[1]{{#1}s\xspace}

\newcommand{\cmv}{{CMV}\xspace}
\newcommand{\Reddit}{Reddit\xspace}
\newcommand{\acomment}{reply\xspace}\newcommand{\comments}{replies\xspace}\newcommand{\Comment}{Reply\xspace} %
\newcommand{\rootcommenter}{root challenger\xspace}
\newcommand{\rootcomment}{root reply\xspace}
\newcommand{\rootcomments}{root replies\xspace}
\newcommand{\subtree}{subtree\xspace}
\newcommand{\commentpath}{path\xspace}
\newcommand{\commentpaths}{\plural{\commentpath}}

\newcommand{\rootedpathunit}{rooted path-unit\xspace}
\newcommand{\commenter}{challenger\xspace}
\newcommand{\commenters}{challengers\xspace}
\newcommand{\commenterapos}{challenger's\xspace}
\newcommand{\fstcomment}{{\em root reply}\xspace}
\newcommand{\fullpath}{{\em full path}\xspace}
\newcommand{\fsttruncated}{{\em root truncated}\xspace}
\newcommand{\resist}{resistant\xspace}
\newcommand{\susc}{malleable\xspace}
\newcommand{\BOW}{\textit{BOW}\xspace}
\newcommand{\POS}{\textit{POS}\xspace}
\newcommand{\addFigure}[2]{\includegraphics[width=#1]{plots/#2}}
\newcommand{\para}[1]{\noindent{\bf #1}}
\newcommand{\figref}[1]{Figure~\ref{#1}}
\newcommand{\secref}[1]{\S\ref{#1}} %
\newcommand{\tableref}[1]{Table~\ref{#1}}
\definecolor{dark-gray}{gray}{0.2}
\newcommand{\nodenum}[1]{\textcolor{dark-gray}{#1}}
\newcommand{\discussiontree}{discussion tree\xspace}
\newcommand{\discussiontrees}{{\discussiontree}s\xspace}
\newcommand{\originalpost}{original post\xspace}
\newcommand{\OP}{OP\xspace}%
\newcommand{\openingargument}{opening argument\xspace}

\newenvironment{fullexample}[1][]
               {
               \list{}{\topsep=2pt\rightmargin=0pt\leftmargin=0pt}%
                \item\relax\small{\textbf{(#1)}%
               }
               \ignorespaces}
               {%
               \endlist
               }

\newcommand{\bigsep}{\noalign{\vskip 2mm}}
\newcommand{\smallsep}{\noalign{\vskip 0.5mm}}

\permission{Copyright is held by the International World Wide Web Conference Committee (IW3C2). IW3C2 reserves the right to provide a hyperlink to the author's site if the Material is used in electronic media.}
\conferenceinfo{WWW 2016,}{April 11--15, 2016, Montr\'eal, Qu\'ebec, Canada.} 
\copyrightetc{ACM \the\acmcopyr}
\crdata{978-1-4503-4143-1/16/04. \\
http://dx.doi.org/10.1145/2872427.2883081}

\begin{document}

\title{
Winning Arguments: Interaction Dynamics and Persuasion Strategies in Good-faith Online Discussions
}

\newcommand{\cucs}{\affaddr{Cornell University}}%
\newcommand{\cuis}{\cucs}%

\numberofauthors{1}
\author{
    \alignauthor
        Chenhao Tan
         \hspace{0.5cm}
        Vlad Niculae
        \hspace{0.5cm}
        Cristian Danescu-Niculescu-Mizil \hspace{0.5cm}
        Lillian Lee\\
    \affaddr{
        Cornell University
        }
\newcommand{\es}{\hspace*{.3in}}
\email{\{chenhao$\vert$vlad$\vert$cristian$\vert$llee\}@cs.cornell.edu}
} %

\hypersetup{pageanchor=false}
\maketitle
\hypersetup{pageanchor=true}

\begin{abstract}
Changing someone's opinion is arguably one of the most important challenges of social interaction.  The underlying process proves difficult to study: it is hard to 
know
 how someone's opinions are formed and whether and how someone's views shift. Fortunately, ChangeMyView, an active community on Reddit, provides a platform where users present their own opinions and reasoning, invite others to contest them, and acknowledge when the ensuing discussions change their original views. In this work, we study these interactions to understand the mechanisms behind persuasion.

We find that persuasive arguments are characterized by interesting patterns of interaction dynamics, such as participant entry-order and degree of back-and-forth exchange.  Furthermore, by comparing similar counterarguments to the same opinion, we
show that language factors play an essential role.  In particular,
the interplay
between the language of the opinion holder and that of the counterargument provides highly predictive cues of persuasiveness.
Finally,
since even in this favorable setting people
may not be persuaded, we
investigate the
problem of determining whether someone's opinion is susceptible to being changed
at all.  For this more difficult task, we show that stylistic choices
in how the opinion is expressed carry
predictive power.
\end{abstract}

\section{Introduction}
\label{sec:intro}

\begin{figure*}
\begin{center}
{\adjincludegraphics[%
width=.95\textwidth,
trim= .0in .1in 0.0in .1in,
clip=true]{%
plots/{ex-fig-tontine-tree-new-label}.png%
}
} %
\caption{\label{fig:intro}
A fragment
of a
``typical''
\fullcmv
\emph{\discussiontree}---typical
in the sense that the full \discussiontree
has an average number of \comments (54),
although we abbreviate or omit many of them for compactness and readability.
Colors indicate distinct users.
Of the 17 \comments shown (in our terminology, every
node except the \originalpost
is
a \acomment),  the \OP explicitly acknowledged only one
as having changed their view:
the starred \acomment
\nodenum{A.1}.
The explicit signal  is the ``$\Delta$'' character in \acomment
\nodenum{A.2}.
(The full \discussiontree is available at \url{https://www.reddit.com/r/changemyview/comments/3mzc6u/cmv_the_tontine_should_be_legalized_and_made_a/}.)
} %
\end{center}
\end{figure*}

Changing a person's opinion is a common goal in
many
settings, ranging from political or marketing campaigns to friendly
or
professional 
conversations.
The importance of this topic has long been acknowledged, leading to
a tremendous amount of research effort
\cite{cialdini1993influence,Dillard:ThePersuasionHandbookDevelopmentsInTheoryAnd:2014,eagly1993psychology,Petty:CommunicationAndPersuasionCentralAndPeripheralRoutes:2012,Popkin:TheReasoningVoterCommunicationAndPersuasion:1994,reardon1991persuasion}.
Thanks to the increasing
number of social interactions online,
\emph{interpersonal persuasion}
has become
observable at a massive scale \cite{fogg2008mass}.
This allows
the study of
interactive
persuasion
{\em in practice, without elicitation},
thus bypassing
some limitations of
laboratory experiments
and leading
to new research questions regarding dynamics in real discussions.
At the same time, the lack of the degree of experimental control offered by lab
trials
raises
new 
methodological
 challenges
that
we address
 in this work.

It is well-recognized that multiple factors are at play in persuasion.
Beyond
({\it i})
the characteristics of the arguments themselves,  such as intensity, valence and
framing
\cite{Althoff+al:14a,bailey2014unresponsive,Bryan:ProceedingsOfTheNationalAcademyOfSciences:2011,burgoon1975toward,hullett2005impact},
and
({\it ii}) social
aspects,
such as social proof and authority
\cite{chaiken1987heuristic,Cialdini01101999,Mitra:ProceedingsOfCscw:2014},
there is also
({\it iii})~the
relationship between the opinion holder and her belief,
such
as her certainty in it and its importance to her
\cite{petty1997attitudes,Pomerantz:JournalOfPersonalityAndSocialPsychology:1995,tormala2002doesn,zuwerink1996attitude}.
Thus,
an ideal setting for the study of persuasion would allow
 access to the
reasoning behind people's
views
in addition to
the
full interactions.
Furthermore,
the outcome of persuasion efforts 
(e.g., which efforts succeed)
should be easy to extract.\footnote{
One might think that the outcome is trivially ``no one ever changes their
mind'', since people can be amazingly resistant to evidence contravening
their beliefs \cite
{CHAMBLISS01071996,youarenotsosmart:11,Nyhan+Riefler:2010a}.
But take heart,  change does
occur, as we shall show.
}

One forum
satisfying these desiderata is
the active \Reddit subcommunity
\fullcmv (henceforth \cmv).%
\footnote{\url{https://reddit.com/r/changemyview}}
In contrast to general
platforms such as Twitter and Facebook,
\cmv requires posters to
state the reasoning
behind
their beliefs %
and %
to reward
successful arguments
with explicit confirmation.
Moreover, discussion quality is
monitored by moderators,
and posters commit to an openness to changing their minds.
The resulting conversations are of reasonably high quality, as demonstrated by
\figref{fig:intro},  showing
the top portion
of
a
\discussiontree
(an original post and all the replies to it)
about legalizing
the ``tontine''.\footnote{
It is not necessary for the reader to be familiar with tontines, but a brief
summary is: a pool of money is maintained where the annual payouts are divided
evenly
among all participants still living.
}
In the figure, \Comment
\nodenum{B.1}
branches off to an
extended back-and-forth
between the blue
original poster (\OP)
and the orange user;
as it turns out, neither ends up yielding, although both remain polite.
\Comment
\nodenum{A.1},
on the other hand, is successful, as the
\OP
acknowledges at
\nodenum{A.2}.
The example suggests that content and phrasing play an important role
(\nodenum{A.1}
does well on both counts), but also that interaction factors
may also correlate with persuasion success. Examples include time of entry
relative to others
and amount of engagement: the discussion at
\nodenum{B.1}
started earlier
than that at
\nodenum{A.1}
and went on for longer.

\para{Outline and highlight reel.}
This work provides three different perspectives on the mechanics of persuasion.
First, we explore how interaction dynamics are
associated with a successful change of someone's opinion~(\secref{sec:exploration}).
We find
(example above to the contrary)
that a \commenter that enters the fray before another tends to have a higher
likelihood of changing the \OP's
opinion;
this holds even for first-time  \cmv \commenters, and so is not a trivial
consequence of more
experienced
disputants contriving to strike first.
Although engaging the OP in
some back-and-forth
is correlated with higher chances of success,
we do not see much \OP conversion in extended conversations.
As for
opinion conversion rates, 
we find that the more  
participants there are in the
effort to persuade the \OP, the larger the likelihood of the \OP changing
her
view; but, interestingly, the relationship is sublinear.

Besides interaction dynamics, language is a powerful tool 
that is
in the full control of 
the
\commenters.
In \secref{sec:comment_pred} we explore this perspective by tackling the 
 task
of predicting which of two {\em similar} counterarguments will succeed in changing the same view. 
By comparing similar arguments we focus on the role of 
stylistic choices
in the presentation of an argument
(identifying reasoning strategies is a separate problem we do not address).
We experiment with style features based solely on the counterargument, as well as with features reflecting the interplay between the counterargument and
the way in which
the view is expressed.
Style features and interplay features both prove useful and outperform a
strong baseline that uses bag-of-words.
In particular,
interplay features alone
 have
  strong predictive power, achieving an improvement of almost 5\% in
accuracy over the baseline method (65.1\% vs 59.6\%) in
a {\em completely fresh}
heldout
dataset.
Our results also show that it is
useful
to include links as evidence---an
interesting contrast to studies of the {\em backfire effect}: ``When
your deepest convictions are challenged by contradictory evidence, your
beliefs get stronger'' \cite{CHAMBLISS01071996,youarenotsosmart:11,Nyhan+Riefler:2010a}.
However,
  it hurts to be
too
intense
in the counterargument.
The feature with the most predictive power of successful persuasion is the
dissimilarity with the \originalpost in word usage, while existing theories
mostly study matching in terms of attitude functions or subject self-%
discrepancy \cite{petty1998matching,tykocinski1994message}.
In the majority of cases, however, opinions are not changed, even
though it takes courage and self-motivation
for the original poster
to post on \cmv and invite other people to change her opinion.
Can we tell whether the OP is unlikely to be persuaded from the way she presents her reasoning? 
In \secref{sec:op_pred}, we turn to this challenging task.
In our pilot study,
humans
found this task quite difficult in a
paired setting and performed
no better than random guessing.
While we can outperform the random baseline in a realistic imbalanced
setting, the AUC score is only 0.54.
Our feature analysis
is consistent with
existing theories on self-affirmation
\cite{cohen2000beliefs,correll2004affirmed} and shows that \susc beliefs are
expressed using more self-confidence and more
organization, in a less
intense
way.
{\em 
While we
 believe that
the
observations we make are
 useful for
understanding persuasion, 
 we do not claim that any of
them have causal explanations.
}

In \secref{sec:discussion}, we discuss
other observations that may open up future directions,
including
attempts to capture higher-level linguistic properties (e.g., semantics and argument structure);
\secref{sec:related} summarizes additional related work and
\secref{sec:conclusion} concludes.
\begin{figure*}[!htb]
\centering
\begin{tabular}{p{0.48\textwidth} p{0.48\textwidth}}
\multicolumn{2}{c}{
\fbox{\parbox{0.98\textwidth}{
\begin{fullexample}[OP]
Title: I believe that you should be allowed to drive at whatever speed you wish as long as you aren't driving recklessly or under extenuating circumstances CMV.\\
I think that if you feel comfortable driving 80 mph or 40 mph you should be
allowed to do so, as long as you aren't in a school or work zone, etc. because
there are a lot more risks in those areas. I think when you're comfortable
driving you will be a better driver, and if you aren't worrying about the
speed limit or cops you are going to be more comfortable. However, I think
that you should only be allowed to drive at whatever speed you wish as long as
you aren't driving recklessly. If you're weaving in and out of traffic at 90,
you probably shouldn't be allowed to go 90, but if you just stay in the fast
lane and pass the occasional person I don't think there is a problem. CMV.
\end{fullexample}
}}} \\ %
\multicolumn{1}{c}{$\downarrow$} & \multicolumn{1}{c}{$\downarrow$}\\
\vspace{-.18in}
\fbox{\parbox[t][][t]{0.48\textwidth}{
\begin{fullexample}[C1]
Some issues with this:
\begin{enumerate}
\item Who's to say what is reckless driving? Where do you draw the line? Speed
is the standard that ensures we know what is considered to be reckless. The idea
of driving any speed you want creates a totally subjective law.
\item How do you judge whether to pass other drives and such? There are a lot
of spatial awareness issues with the roads being so unpredictable.
\item How do you expect insurance and courts to work out who's at fault
for an accident?
\end{enumerate}
A: ``Yeah this guy was going 100 mph!''

B: ``But I wasn't driving recklessly - you were!''

It's simply not realistic and creates some serious legal issues.
\end{fullexample}
}} %
&
\vspace{-.18in}
\fbox{\parbox[t][][t]{0.48\textwidth}{
\begin{fullexample}[C2]
They're many issues I have with this idea but I'll start with the most
pressing one. Think of the amount of drivers you pass by every day. Imagine
all of them going at whatever speed they choose. How would this work? You
cannot have a driver going 35 and a driver who wants to go 65 in the same
lane.

Now lets take this onto the highway and you can see how horrific this could
get quickly. They're too many drivers out on the road for everyone to choose
there own speed.

Speed limits protect us all because it gives us a reasonable expectation in
whatever area we're driving in. Have you ever been on the highway being a
driver going 40mph? If you're doing the speed limit (65) you catch up to them
so fast you barely have time to react before an accident occurs. You aren't
expecting this low speed when everyone is going at similar speeds to yours.

Drivers need to know the speed expectations so they can drive and react
accordingly. If everyone goes at whatever speed they want it will only cause
many many accidents.
\end{fullexample}
}} %
\end{tabular}
\caption{\label{tab:pair}
An
\emph{\originalpost} and a pair of \emph{\rootcomments} \textbf{C1} and \textbf{C2} contesting it,
where \textbf{C1} and \textbf{C2} have relatively high vocabulary overlap with each other,
but only one changed the OP's opinion.  (\secref{sec:comment_pred} reveals which one.)}
\end{figure*}

\section{Dataset}
\label{sec:data}

\newcommand{\submission}{\originalpost}
We draw our data from
the
\fullcmv subreddit
(\cmv), which has over 211,000 subscribers to date.
It is self-described\footnote{Quotations here are
from the \cmv 
wiki.} as
``dedicated to the civil
discourse
[sic] of opinions''.
\cmv is well-suited to our purposes because of
its setup and mechanics, the high quality of its arguments, and the size and
activity of its user base. We elaborate below.

The mechanics of the site are as follows.
Users that ``accept that they may be wrong or want help
changing their view''
submit
\emph{{\submission}s}, and readers are invited
to argue
for the other side. The original posters (\emph{{\OP}s})
explicitly recognize
arguments that succeed in changing their view by replying with the
{\em delta} ($\Delta$) character
(an example is node 
\nodenum{A.2} in Figure
\ref
{fig:intro}) and including ``an explanation as to why and how'' their view
changed.
A {\Reddit} bot
called the DeltaBot
confirms deltas (an example is 
\nodenum{A.3} 
in Figure \ref{fig:intro}) and
maintains
a {leaderboard}
of per-user 
$\Delta$
counts.\footnote{%
	Although non-OPs can also issue deltas, in this work, we only count
	deltas given by a user in their OP role.  A consequence is that we only
	consider
	{\discussiontree}s where the OP's
	\Reddit
	account had not been deleted---i.e., the
	\submission
	is not attributed to the ambiguous name ``[deleted]''--- at the time of
	crawl.
}
The experimental advantages of this setup include: \\
(1) Multiple users make different attempts at changing the same
person's mind on the same issue based on the same rationale,
thus controlling for a number of variables but providing variation
along other important aspects.
\figref{tab:pair}, for example,
presents in full
two counter-arguments, C1 and C2.
They both respond to the same claims,   but
differ in style, structure,
tone, and other respects. \\
(2) The deltas
serve as explicit persuasion
labels that are (2a) provided by the actual participants and (2b) at the
fine-grained level of individual arguments, as opposed to mere indications that
the OP's view was changed.
\\
(3) The OP has, in principle,
expressed an openness to other points of view, so that we might hope to extract
a sufficient number of view-changing examples. \\
These advantages are not
jointly
exhibited by other debate sites, such as 
CreateDebate.com, ForandAgainst.com, or Debate.org.

The high quality of argumentation makes \cmv a model site for seeing whether
opinion shifts can at least occur  under favorable conditions.
Moderators enforce {\cmv} rules, making sure that OPs
{explain} why they
hold their beliefs
and do so
at reasonable length (500 
characters
or more),
and
that OPs
engage in
conversation with
challengers
in a timely fashion.
Other rules apply to those who contest the \submission.
There are rules intended to prevent ``low effort'' posts,
 such as ``Posts that are only a
single link with no substantial argumentation'', but ``Length/conciseness isn't
the determining [criterion]. Adequate on-topic information is.''\footnote{It is worth noting that, as in many online
communities, not all these rules were
in place
at the site's creation.
It is a
separate and interesting research question to understand what effects these
rules have and why they were put in place.
The currently enforced set of rules
is available at \url{https://www.reddit.com/r/changemyview/wiki/rules}.}
\figref{tab:pair} shows an example where indeed, the OP described their
point in
reasonable detail, and the responders raised sensible objections.%
%
%
%
%
%
%
%

%
%

%
%
%

\begin{figure}[t]
\centering
	\begin{subfigure}[t]{0.23\textwidth}
        \addFigure{\textwidth}{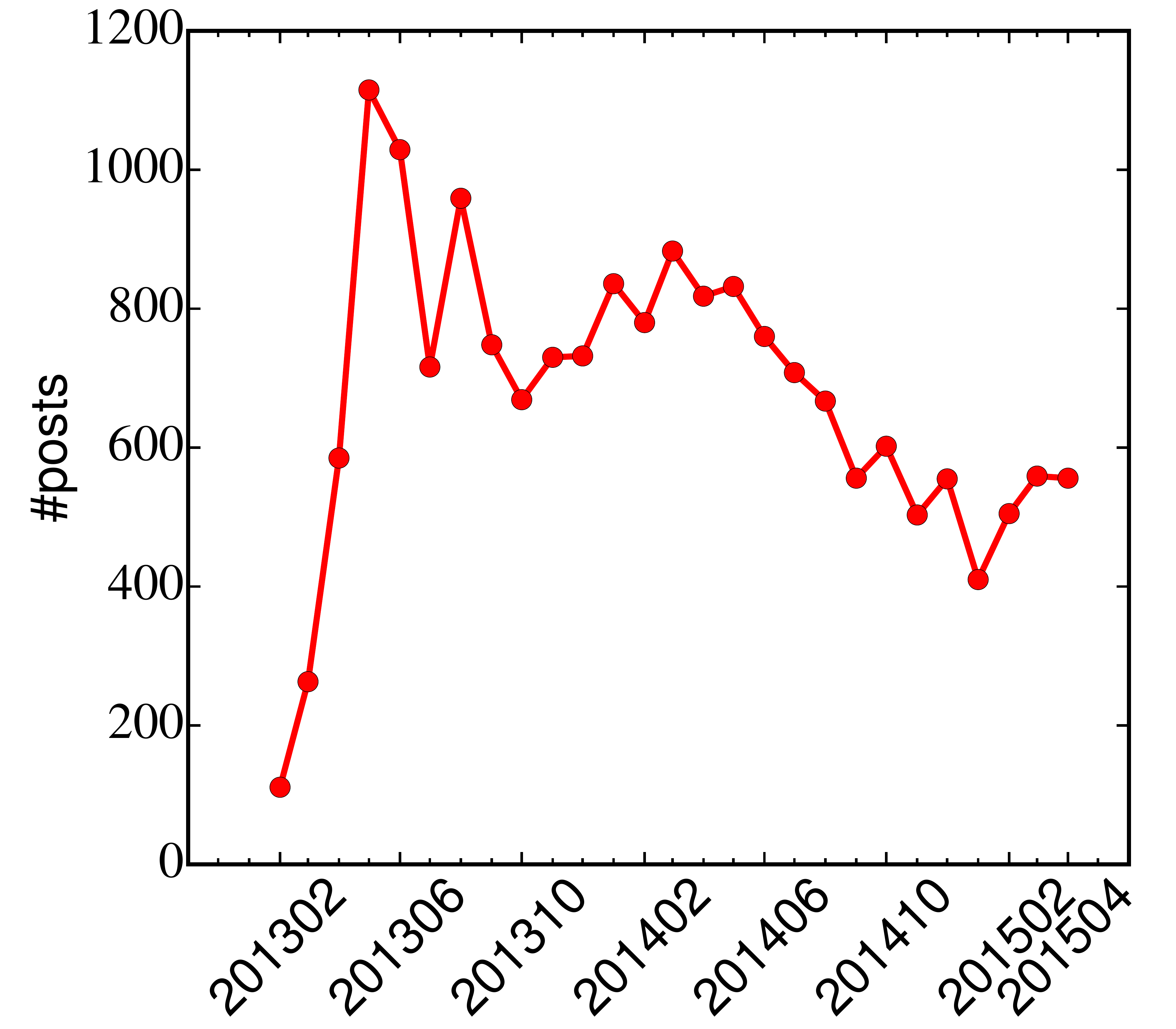}
        \caption{Number of posts per month.}
        \label{fig:num_posts}
    \end{subfigure}
    \hfill
	\begin{subfigure}[t]{0.23\textwidth}
		\addFigure{\textwidth}{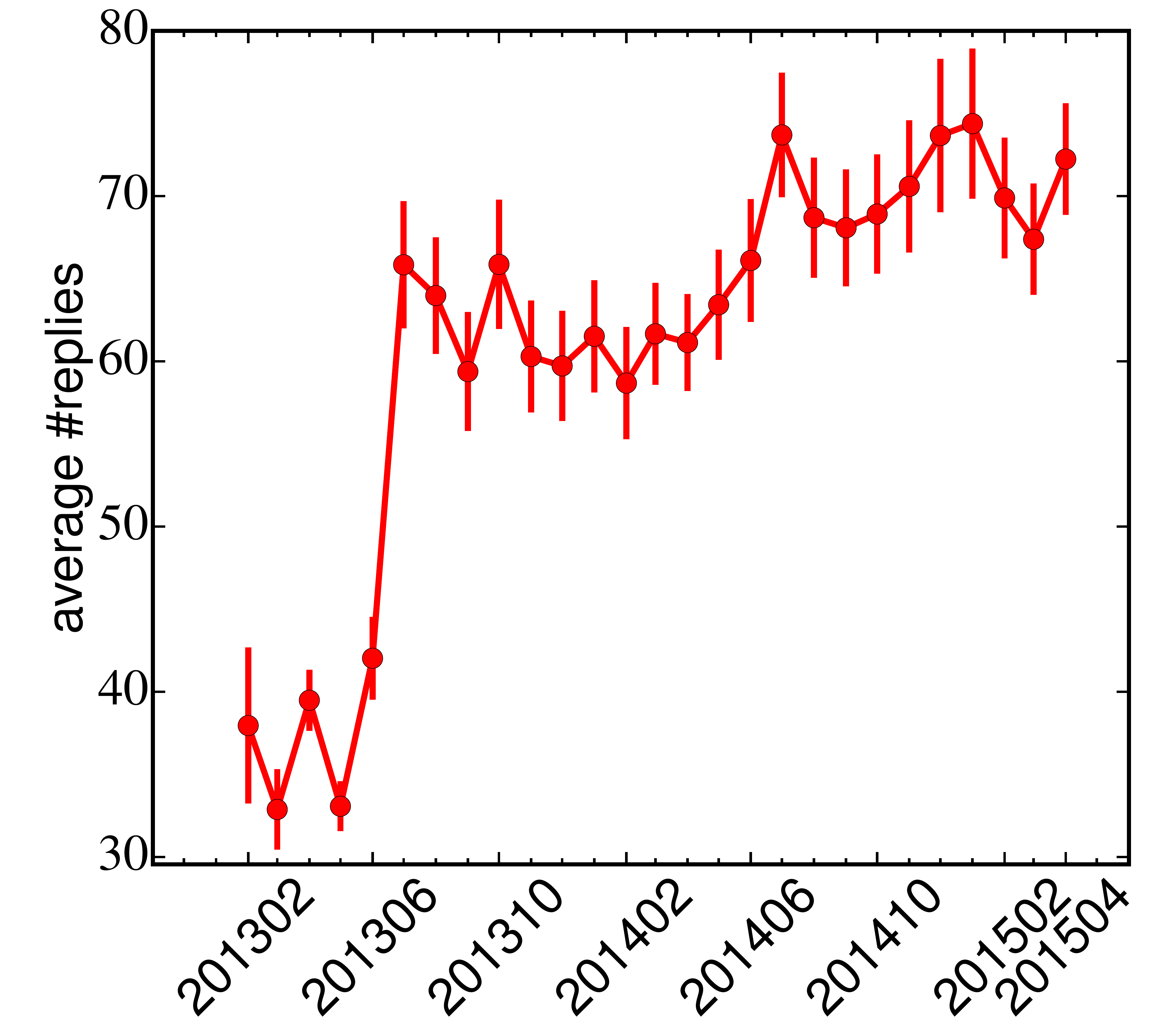}
		\caption{Average no. \comments per post.}
		\label{fig:avg_comment}
	\end{subfigure}

	\begin{subfigure}[t]{0.23\textwidth}
        \addFigure{\textwidth}{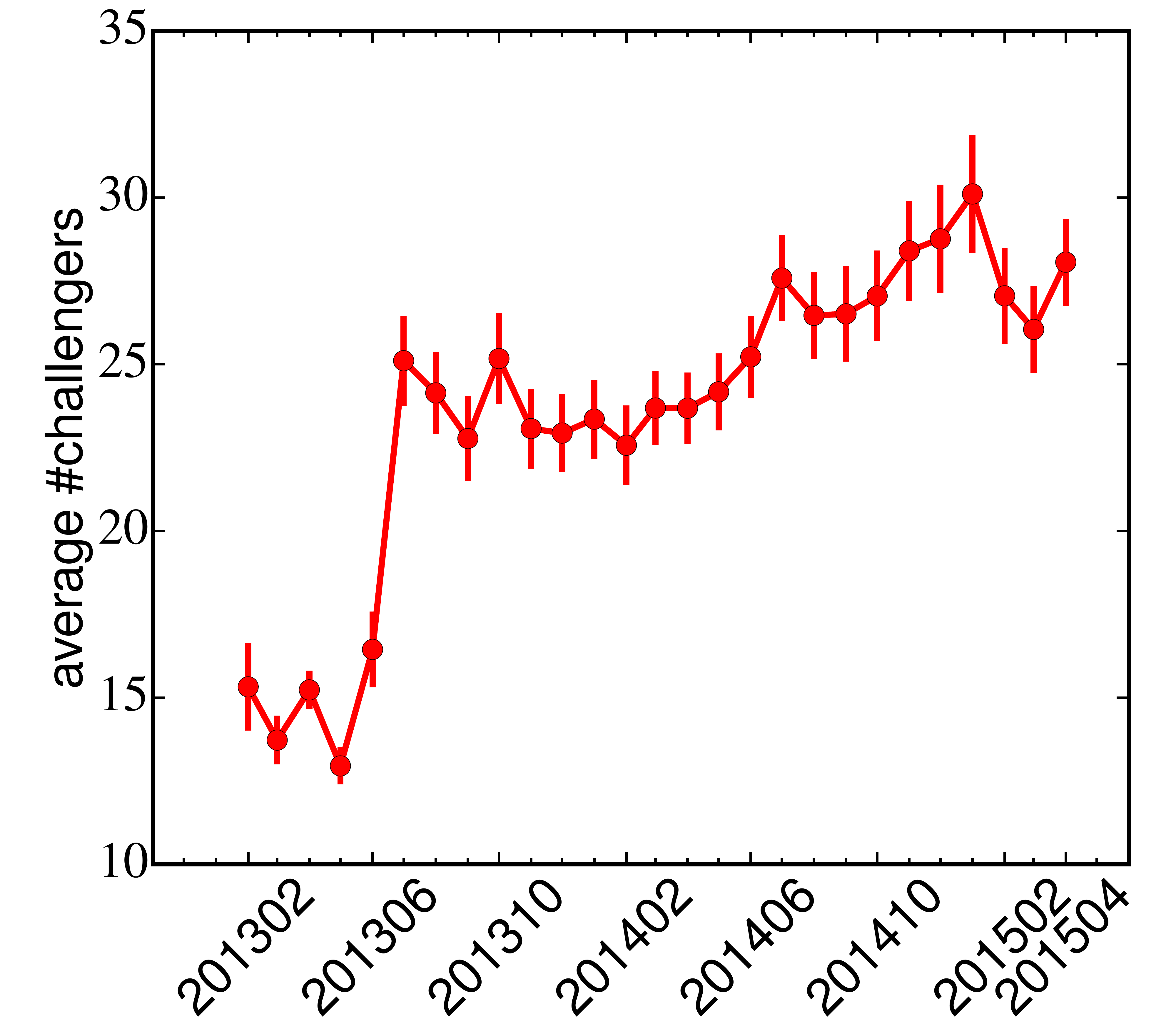}
        \caption{Average no. \commenters per post.}
        \label{fig:avg_commenter}
    \end{subfigure}
    \hfill
	\begin{subfigure}[t]{0.23\textwidth}
		\addFigure{\textwidth}{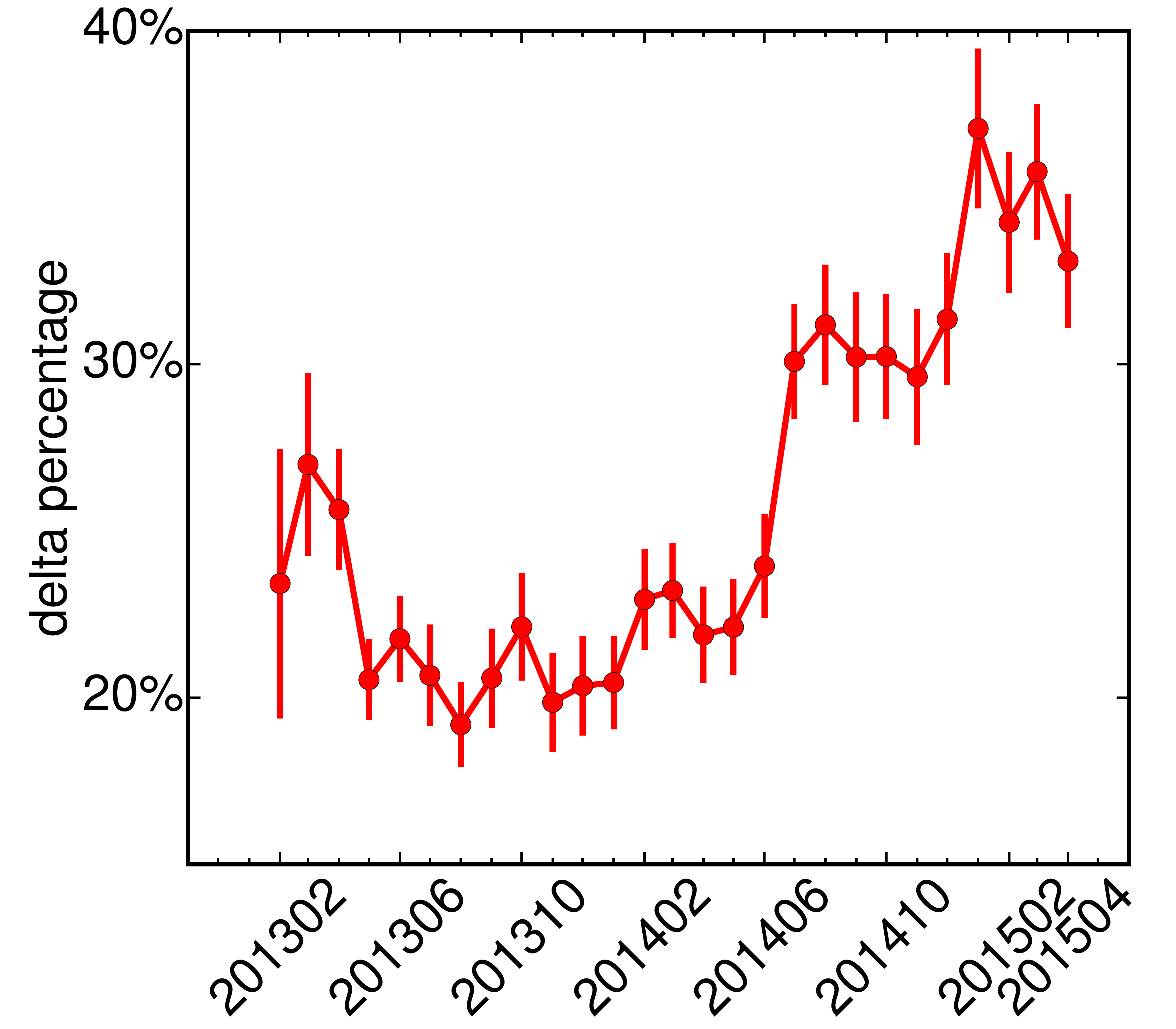}
		\caption{Average delta percentage.}
		\label{fig:avg_delta}
	\end{subfigure}
	\caption{Monthly activity
	over
	all full months represented in
	the training set. The {\em delta percentage} is the fraction of {\discussiontree}s
	in which the OP awarded a delta.
	\label{fig:health}}
\end{figure}

%
% At the time of writing, {\cmv} has over 207\,000 subscribers, .
The high amount of activity on \cmv means that we can extract a large amount of
data.
We
process
all {\discussiontree}s
created at any time from January 2013, when the
subreddit
was
created, to August 2015, saving roughly the final 4 months (May--August
2015) for held-out evaluation.
Some size statistics are given in \tableref{tab:data-stats}.
Monthly trends are depicted in \figref{fig:health}:\footnote{We omit the first month as the DeltaBot may not have been set up.}
after the initial startup, activity levels stabilize
to a healthy, stable growth in average number of \comments and \commenters,
as, gratifyingly, do \OP 
conversion
rates, computed as the
fraction
of {\discussiontree}s wherein the OP awarded a 
$\Delta$
(\figref{fig:avg_delta}).
For posts where the \OP gave at least one delta, the OP gave 1.5 deltas on average.
This dataset 
is
available at \url{https://chenhaot.com/pages/changemyview.html}.

\begin{table}[t]
\caption{%
Dataset statistics. The disjoint training and test date ranges are
2013/01/01--2015/05/07
and 2015/05/08--2015/09/01.
}

\centering
\small
\begin{tabular}{l r r r r}
\toprule
& \# {\discussiontree}s & \#
nodes
& \# OPs & \#
uniq.
participants
\\
\midrule
Training %
& 18,363 & 1,114,533 & 12,351 & 69,965 \\
Heldout %
& 2,263 & 145,733 & 1,823 & 16,923 \\
\bottomrule
\end{tabular}
\label{tab:data-stats}
\end{table}

\section{Interaction dynamics}

\label{sec:exploration}

Changing someone's opinion is a complex process, often involving repeated interactions between the participants.  In this section we investigate the relation between the underlying dynamics and the chances of
``success'',
where ``success'' can be seen from the perspective of the \commenter (did she succeed in changing the {\OP}'s opinion?), as well as from that of
the set of \commenters
(did anyone change the {\OP}'s view?).

In order to discuss the relation between interaction dynamics and success, we now introduce corresponding terminology
using the example illustrated in \figref{fig:intro}:

\begin{itemize}
\item An original statement of views ({\em \originalpost}) together with all the replies form a {\em \discussiontree}.
\item A direct reply to an \originalpost is called
a
 {\em \rootcomment}
(\nodenum{A.1} and \nodenum{B.1}
 in \figref{fig:intro}). The author of a \rootcomment is a {\em \rootcommenter}.
\item  A {\em \subtree} includes a \rootcomment and all its children
(\nodenum{B.1--B.12}
form one of the two subtrees in \figref{fig:intro}).
\item A {\em \commentpath} constitutes all nodes from \rootcomment to a leaf node.  \figref{fig:intro} contains four \commentpaths: $P_1$:
\nodenum{A.1},
$P_2$:
\nodenum{A.1, A.2, A.4, A.5},
$P_3$:
\nodenum{B.1--B.11}
and $P_4$:
\nodenum{B.1, B.12}.
Note that when a $\Delta$ is awarded, the DeltaBot automatic reply
(\nodenum{A.3})
and the OP's post that triggers it
(\nodenum{A.2})
are not considered part of the \commentpath.
\end{itemize}

In order to focus on discussions with
non-trivial
activity, in this section we only consider \discussiontrees with at least 10 replies from \commenters and at least one reply from the \OP.

\subsection{Challenger's success}

A \commenter is successful if
she
manages to change the view of the OP and receive a $\Delta$.  We now examine how the interaction patterns in a \discussiontree relate to a \commenter's success.

\begin{figure}[t]
\centering
	\begin{subfigure}[t]{0.23\textwidth}
		\addFigure{\textwidth}{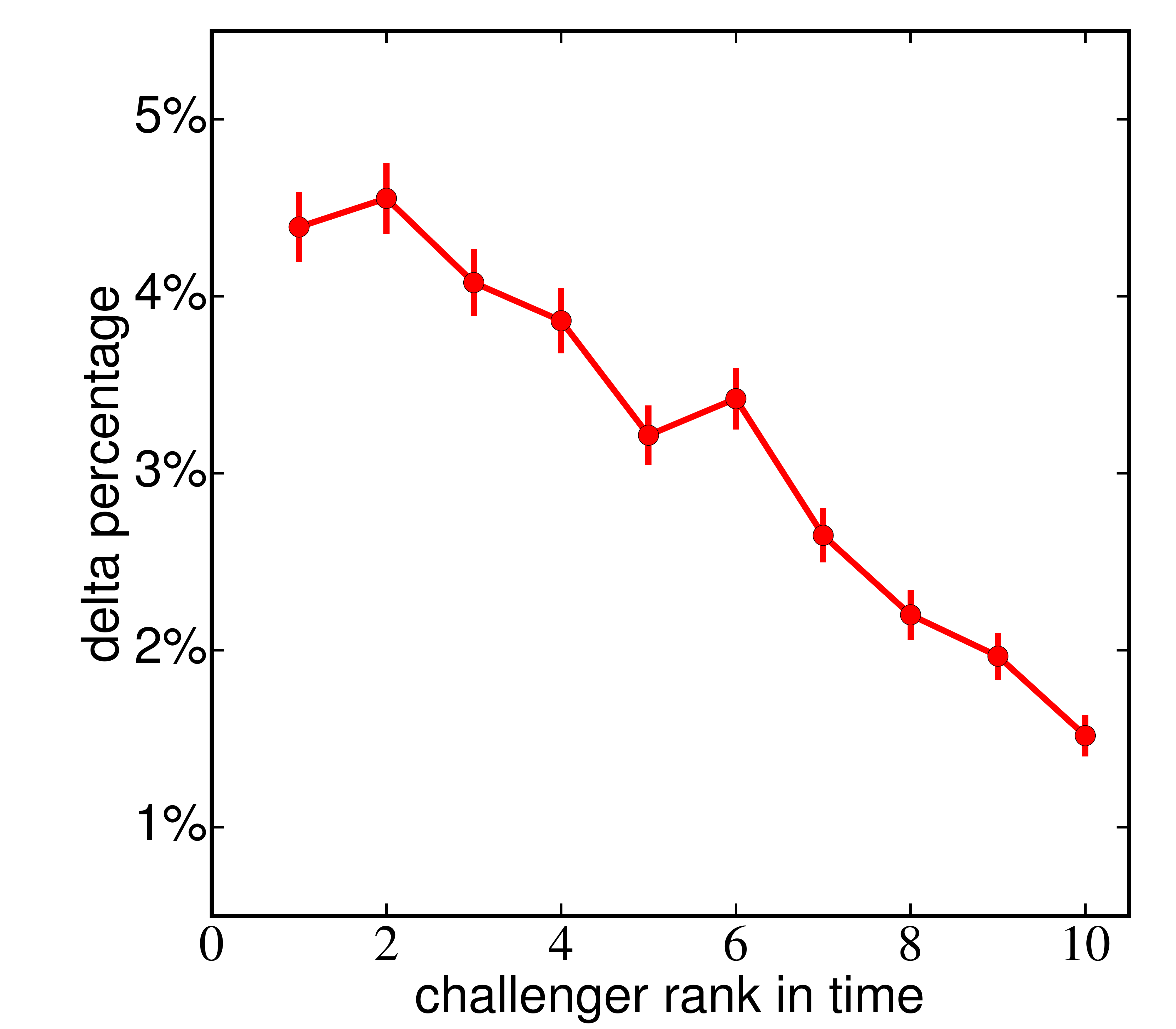}
		\caption{Delta ratio vs. entry order.}
		\label{fig:entry_order_all}
	\end{subfigure}
	\hfill
	\begin{subfigure}[t]{0.23\textwidth}
		\addFigure{\textwidth}{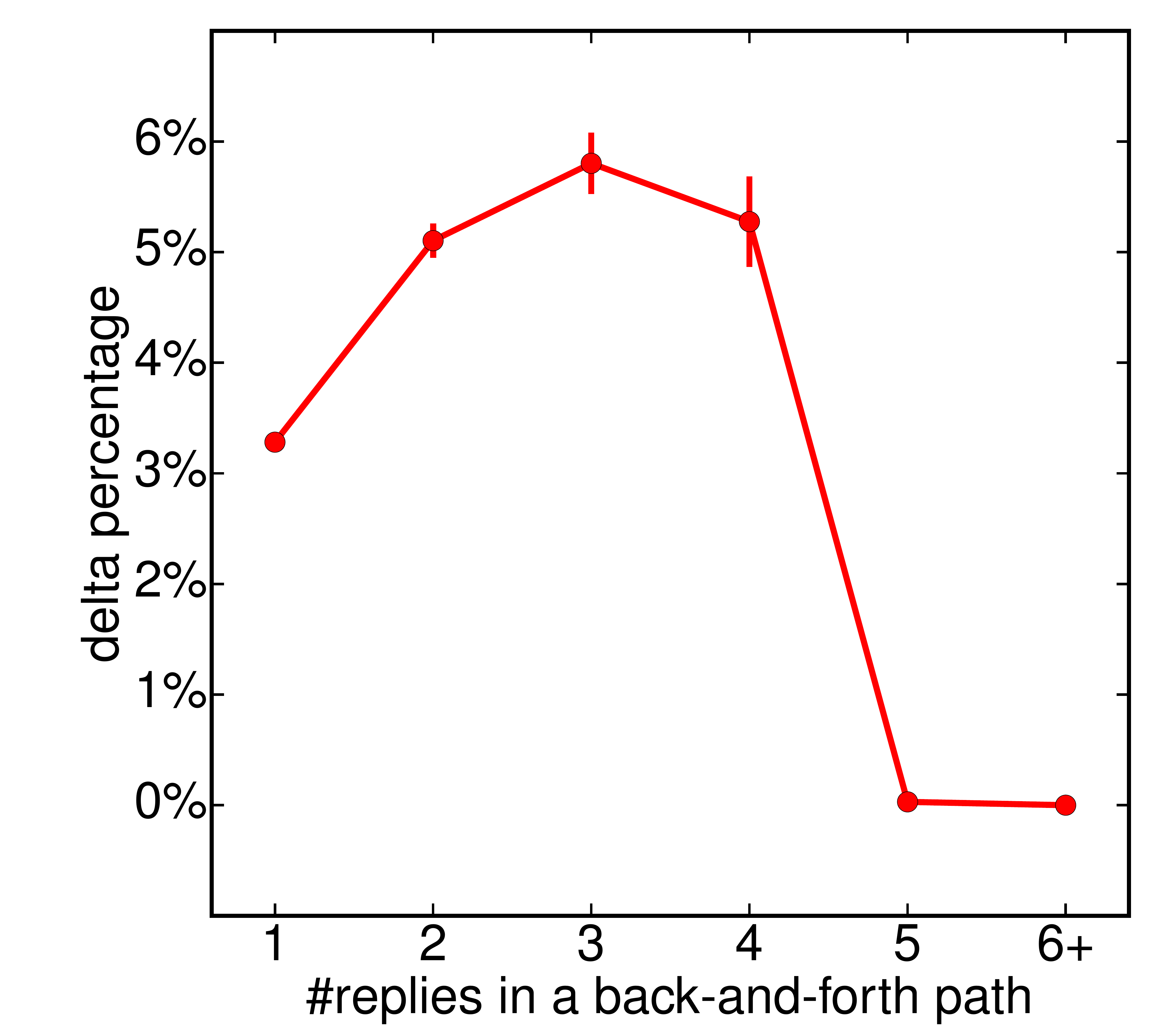}
		\caption{Delta ratio vs. degree of back-and-forth exchanges.}
		\label{fig:back_forth}
	\end{subfigure}
	\caption{
	\figref{fig:entry_order_all} shows
	the ratio of a person
	eventually winning a delta
	in a post with at least 10 \commenters depending on the order of her/his entry.
	{\em Early entry is more likely to win a delta.}
	\figref{fig:back_forth} presents the probability of winning a delta given the number of comments by a \commenter
	in a back-and-forth path with OP.
	With 
	6 or more replies in a back-and-forth path, {\em no} \commenters managed to win a delta among our 129 data points (with 5 replies, the success ratio is 1 out of 3K).
	In both figures, error bars represent standard errors (sometimes 0).
	\label{fig:entry_order}}
\end{figure}

\para{Entry time.}
How does the time when a \commenter enters a discussion
relate
to her chances of success?  A late entry might give the \commenter time to read
attempts
by
other \commenters
and better formulate their
arguments, while an early entry might give her the
first-mover advantage.%
\footnote{Note that although
reply display order
is affected by upvotes, entry time is an important factor when the \OP follows the post closely.}
Even for
{\originalpost}s
that eventually attract attempts by 10 unique \commenters, the first two \commenters are 3 times
more
likely to succeed as the 10\textsuperscript{th}
(\figref{fig:entry_order_all}).

One
potential
explanation
for this finding
is that dedicated expert users are more likely to be more active on the site and thus
see posts first.
To account for this, we redo the analysis only for users that are participating for the first time on \cmv.
We observe that even after controlling for user experience,
an earlier entry time is still more favorable.
We omit the figure for space reasons.

\para{Back-and-forth.}
After entering a discussion, the \commenter can either spend more effort and engage with the OP in a back-and-forth type of interaction or call it quits.
\figref{fig:back_forth} shows the relation between the likelihood of receiving a $\Delta$ and degree of back-and-forth, defined as the number of \comments the \rootcommenter made in a \commentpath
involving only her and the OP.%
\footnote{%
If a subtree won a $\Delta$, we only consider the winning path; otherwise,
other conversations would be mistakenly labeled unsuccessful.
For instance, the path
\nodenum{A.1, A.2, A.4, A.5}
in \figref{fig:intro} is not considered.
}
We observe a non-monotonic relation between back-and-forth engagement and likelihood of success:
perhaps
while some engagement signals the interest of the OP, too much engagement can indicate futile insistence; in fact, after 5 rounds of back-and-forth the \commenter has virtually no
chance
of receiving a $\Delta$.

\subsection{OP's conversion}
\label{sec:exploration-op}  %

From the perspective of an
\originalpost,
conversion can happen when any of the \commenters participating in the discussion succeeds in changing
the \OP's view.
We now turn to exploring how
an \OP's conversion
relates to the volume and type of activity
  her \originalpost attracts.

\begin{figure}[t]
\centering
	\begin{subfigure}[t]{0.23\textwidth}
		\addFigure{\textwidth}{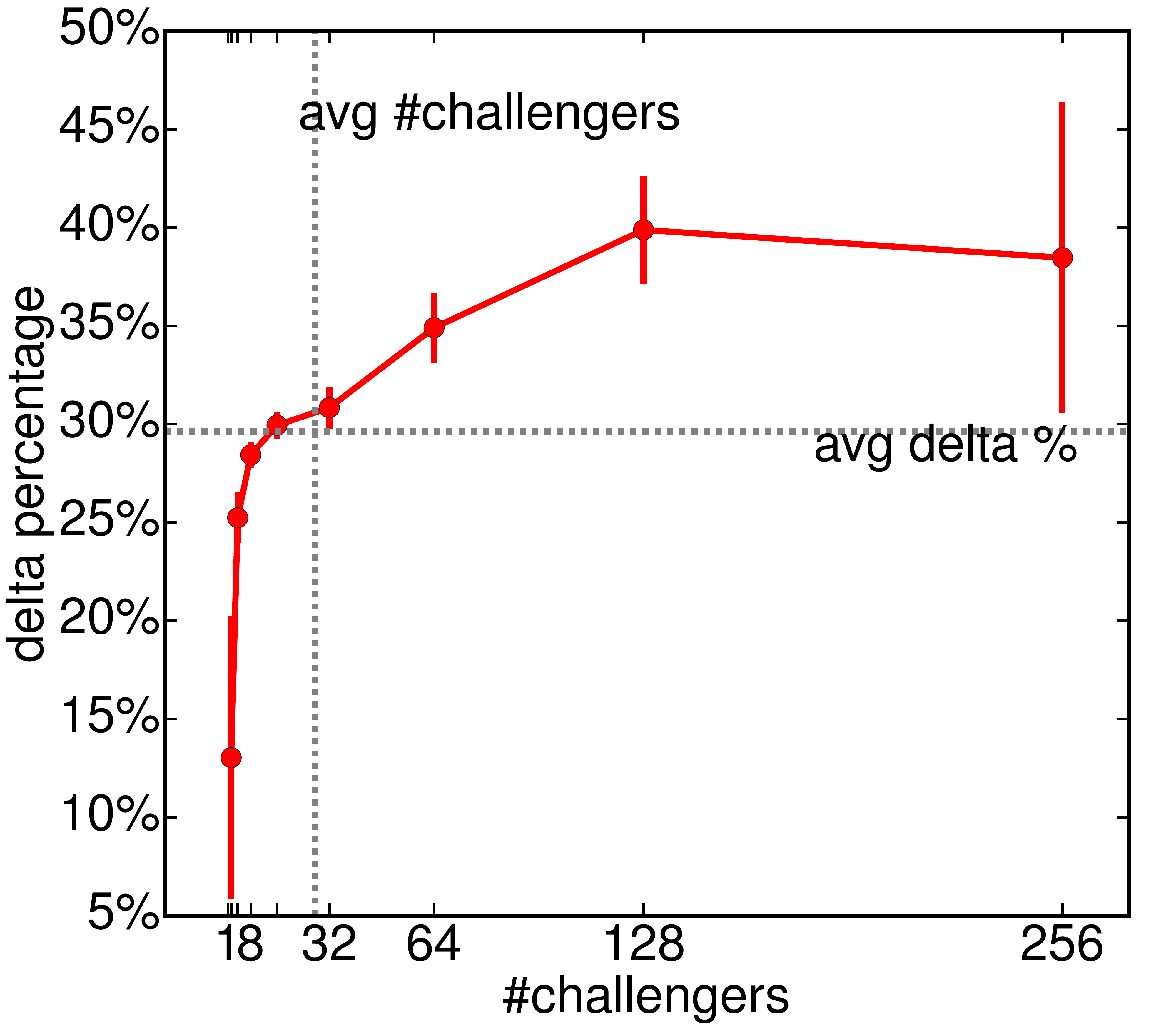}
		\caption{Delta percentage vs. number of unique \commenters.}
		\label{fig:delta_op_commenter}
	\end{subfigure}
	\hfill
	\begin{subfigure}[t]{0.23\textwidth}
		\addFigure{\textwidth}{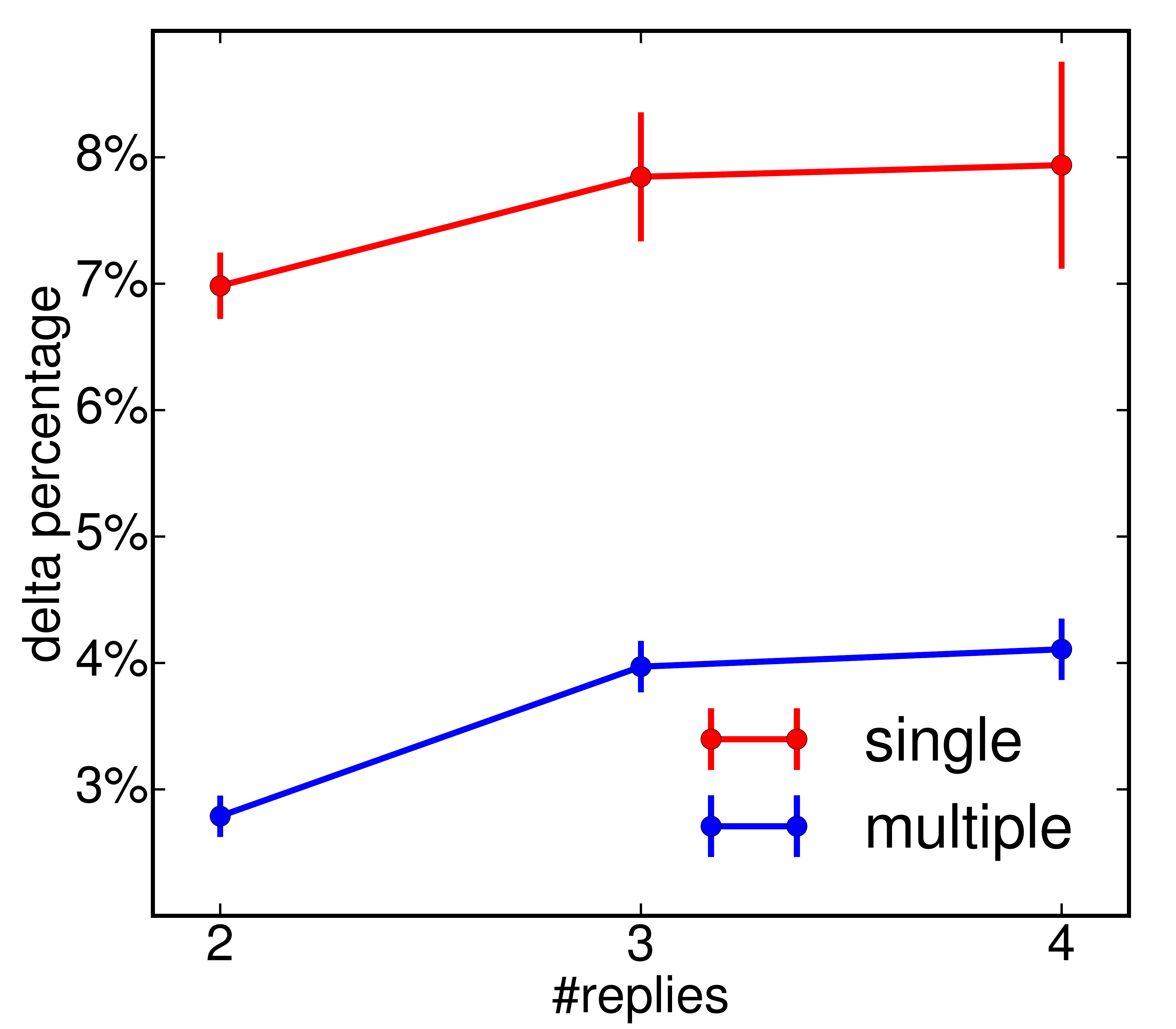}
		\caption{Single-\commenter subtree vs. multiple-\commenter subtree controlled by the number of \comments.}
		\label{fig:delta_subtree_commenter}
	\end{subfigure}
	\caption{%
	Probability that a submitted view will be changed, given (a) the total number
	of unique \commenters binned using $\log_2$, and (b) the number of \comments in a subtree.
	\label{fig:delta_op}}
\end{figure}

\para{Number of participants.}  It is reasonable to expect that
an OP's conversion
is tied to the number of \commenters
\cite{chaiken1987heuristic,Cialdini01101999}.
For instance, the OP might be persuaded by observing the sheer number of people arguing against her original opinion.
Moreover, a large number of \commenters will translate
into
a more diverse set of arguments, and thus higher
likelihood
that the OP will encounter the ones that
best fit
her situation.
Indeed, \figref{fig:delta_op_commenter} shows
that the likelihood of conversion does increase with the number of unique \commenters.
Notably,
we observe a saturation
in
how much value each new \commenter adds beyond a certain point.
\para{Sheer number of \commenters or diversity of counterarguments?}
\noindent
 To distinguish between the two possible explanations proposed in the previous paragraph,
we control for the diversity of counterarguments by focusing only on subtrees,
in which \commenters generally focus on the same argument.
To make a fair comparison,
we further control the number of total \comments in the \subtree.
In light of \figref{fig:back_forth}, we only consider
{\subtree}s with between 2 and 4 \comments.
\figref{fig:delta_subtree_commenter} shows that
single-\commenter subtrees consistently outperform multiple-\commenter subtrees
in terms of conversion rate.
This observation suggests that
the sheer number of \commenters
is not necessarily associated with
 higher chances of conversion.
 The fact that multiple-\commenter subtrees are less effective
 might suggest that
 when talking about the same counterargument, \commenters might not be adding value to it,
 or they might even disagree (e.g.,
\nodenum{B.12} vs.~\nodenum{B.2} in \figref{fig:intro}); alternatively, \rootcomments that attract multiple \commenters might be less effective to begin with.

\section{
Language indicators
\\ of persuasive arguments}
\label{sec:comment_pred}

The
interaction
dynamics studied in the previous section are
to a large extent
outside the \commenterapos\ influence. The language used in arguing,
however, is under one's complete control;
linguistic
correlates
of successful
persuasion can therefore prove of practical value
to aspiring persuaders.
In order to understand what
factors of language are effective, we set up paired prediction tasks to
explore
the effectiveness of
textual discussion features,
in the
context of \cmv.

\subsection{Problem setup}
In order to study an individual's success in persuasion, we consider the collection of arguments from the same person in the same line of argument.
We focus on arguments from {\rootcommenter}s since the \rootcomment is what initiates a line of argument and determines whether the OP will choose to engage.
We define all \comments in a \commentpath by the \rootcommenter as a {\em \rootedpathunit}, e.g., \acomment
\nodenum{A.1} and \nodenum{B.1}
in \figref{fig:intro}.

As shown in \secref{sec:exploration},
situations
where
there
is more than one \acomment
in a \rootedpathunit
correspond to a higher chance that
the OP will be persuaded.
So, while the \commenter's \openingargument should be important, statements
made later in the \rootedpathunit could be more important.
To distinguish these two cases, we consider
two
related
prediction tasks:
\fstcomment, which only uses the \commenter's \openingargument in a \rootedpathunit,
and \fullpath, which considers the text in all \comments within a \rootedpathunit.
In response to
the same \originalpost, there are many possible ways to change
someone's
view. We aim to
find linguistic factors that can help one formulate her/his argument,
rather than to analyze
reasoning strategies.%
\footnote{%
That
is an intriguing
problem 
for future work
that requires a knowledge base and sophisticated semantic
understanding of language.} Hence, for each \rootedpathunit that wins a $\Delta$, we
find the
\rootedpathunit
in the same \discussiontree
that did not win a $\Delta$
but was the most ``similar'' in
topic.
We
measure
similarity between {\rootedpathunit}s based on Jaccard similarity
in the
\rootcomments
after removing stopwords
(as defined by Mallet's dictionary \cite{McCallumMALLET}).

$$\operatorname{Jaccard}(A, B) = \frac{|A \cap B|}{|A \cup B|},$$
\noindent where
$A$, $B$
are
the sets of words in the first \acomment of
each of
the two {\rootedpathunit}s. This leads to a
balanced binary prediction task: which of the two lexically similar
{\rootedpathunit}s is the successful one?  With this setup, we attempt to roughly de-%
emphasize {\em what} is being said, in favor of {\em how} it
is
expressed.

We further avoid trivial cases, such as
replies that are not arguments but
clarifying questions, by removing
cases
where the
\rootcomment
has fewer than 50 words.
In order to make
sure that there are enough counterarguments that
the
OP saw,
motivated by the results in \secref{sec:exploration-op},
we also require that
there are
at least
10
\commenters in the \discussiontree and at least 3 unsuccessful {\rootedpathunit}s before the last \acomment that
the
OP made in the \discussiontree.
In an ideal world, we would control for both
length \cite{Danescu-Niculescu-Mizil+Cheng+Kleinberg+Lee:12} and topic
\cite{jaech-EtAl:2015:EMNLP,tan+lee+pang:14},
but we don't have the luxury of having enough data to do so.
In our pilot experiments,
annotators
find that Jaccard-controlled pairs
are easier to compare
than length-matched pairs, as
the lexical control is likely to produce arguments that make similar claims.
Since
length can be predictive
(for instance, {\bf C2} won a $\Delta$ in \figref{tab:pair}),
this raises the concern of false positive findings.
Hence we develop a
post-mortem
``dissection'' task (labelled \fsttruncated)
in which we only consider the \rootcomment and truncate the longer one within a pair so that both \rootcomments
have
the same number of words.
This
forcefully removes all length effects.

\para{Disclaimer:} %
{
Features that lose predictive power
in the \fsttruncated setting
  (or ``reverse direction''\footnote{E.g., more of feature $f$ is significantly better for \fstcomment, but less $f$ is significantly better in
\fsttruncated.
})
are
not necessarily false positives (or non-significant),
as truncation
can remove
significant
fractions of the text and lead to different distributions in the resultant dataset.
Our point, though, is: if features retain predictive power \emph{even in} the root truncated settings, they
must be
indicative beyond length.
}

We extract pairs
from the training and heldout periods
respectively as
training data (3,456 pairs) and heldout testing data (807 pairs).
Given that our focus is on language, we only use text-based features in this section.\footnote{%
An entry order baseline only achieves 54.3\% training accuracy.
}
In preprocessing, we remove
explicit edits that users made after
posting or commenting,
and convert quotations and URLs into special tokens.

\subsection{Features}
\label{sec:pair_features}

In order to capture characteristics of successful
arguments,
we explore two classes of
textual
features:
(\secref{sec:pair_simi_features}) features that describe the interplay between
a particular \commenter's 
\comments
and the {\originalpost},
and 
(\secref{sec:pair_comment_features}) features that are solely based on
his/her
\comments.
We present those
features that
are
statistically significant
in the training data under the paired t-test with Bonferroni
correction %
for multiple comparisons.

\begin{table}[t]
\centering
\caption{Significance
tests on interplay features.\label{tb:pair_simi_test}
Features are sorted by average p-value in the two tasks.
In all
feature testing tables,
the number of arrows indicates the level of
p-value, while the direction shows the relative relationship between positive instances and negative instances, 
$\uparrow\uparrow\uparrow\uparrow$: $p<0.0001$, $\uparrow\uparrow\uparrow$: $p<0.001$, $\uparrow\uparrow$: $p<0.01$,
$\uparrow$: $p<0.05$. $T$ in the \fstcomment column indicates that the feature is also significant in
the \fsttruncated condition,
while $T^R$ means that it is
significant
in \fsttruncated
but the direction is reversed.
}
\small
\begin{tabular}{p{1.5in}ll}
\toprule
Feature name & \fstcomment & \fullpath\\
\midrule
\acomment frac. in all & \makebox[17pt][l]{$\downarrow\downarrow\downarrow\downarrow$} \textcolor{gray}{($T$)} & $\downarrow\downarrow\downarrow\downarrow$ \\
\acomment frac. in content & \makebox[17pt][l]{$\downarrow\downarrow\downarrow\downarrow$} \textcolor{gray}{($T$)} & $\downarrow\downarrow\downarrow\downarrow$ \\
\OP frac. in stopwords & \makebox[17pt][l]{$\uparrow\uparrow\uparrow\uparrow$} \textcolor{gray}{($T^R$)} & $\uparrow\uparrow\uparrow\uparrow$ \\
\#common in stopwords & \makebox[17pt][l]{$\uparrow\uparrow\uparrow\uparrow$} \textcolor{gray}{($T^R$)} & $\uparrow\uparrow\uparrow\uparrow$ \\
\acomment frac. in stopwords & \makebox[17pt][l]{$\downarrow\downarrow\downarrow\downarrow$}  & $\downarrow\downarrow\downarrow\downarrow$ \\
\OP frac. in all & \makebox[17pt][l]{$\uparrow\uparrow\uparrow\uparrow$} \textcolor{gray}{($T^R$)} & $\uparrow\uparrow\uparrow\uparrow$ \\
\#common in all & \makebox[17pt][l]{$\uparrow\uparrow\uparrow\uparrow$} \textcolor{gray}{($T^R$)} & $\uparrow\uparrow\uparrow\uparrow$ \\
Jaccard in content & \makebox[17pt][l]{$\downarrow\downarrow\downarrow\downarrow$} \textcolor{gray}{($T$)} & $\downarrow\downarrow\downarrow\downarrow$ \\
Jaccard in stopwords & \makebox[17pt][l]{$\uparrow\uparrow\uparrow\uparrow$} \textcolor{gray}{($T^R$)} & $\uparrow\uparrow\uparrow\uparrow$ \\
\#common in content & \makebox[17pt][l]{$\uparrow\uparrow\uparrow\uparrow$} \textcolor{gray}{($T^R$)} & $\uparrow\uparrow\uparrow\uparrow$ \\
\OP frac. in content & \makebox[17pt][l]{$\uparrow$} \textcolor{gray}{($T^R$)} & $\uparrow\uparrow\uparrow\uparrow$ \\
Jaccard in all & \makebox[17pt][l]{$\downarrow$} \textcolor{gray}{($T$)} &  \\

\bottomrule
\end{tabular}
\end{table}

\subsubsection{Interplay with
the
\originalpost: \tableref{tb:pair_simi_test}}
\label{sec:pair_simi_features}

The context established by the
{\OP}'s
statement of her view can
provide valuable information in judging the relative quality of
a \commenter's arguments.
We capture the interplay between arguments and 
{\originalpost}s
through
similarity metrics based on word overlap.\footnote{We also tried
\textit{tf-idf}, topical, and word embedding--based similarity in cross
validation on training data. We defer
discussion of
potentially useful
features 
to
 \secref{sec:discussion}.} We consider four variants based on the
number of unique
words in common between
the argument ($A$) and
the
\originalpost ($O$):

\begin{itemize}[itemsep=0pt]
\item number of common words:
$|A \cap O|$,
\item \acomment fraction:
$\frac{|A \cap O|}{|A|}$,
\item \OP fraction:
$\frac{|A \cap O|}{|O|}$,
\item Jaccard:
$\frac{|A \cap O|}{|A \cup O|}$.
\end{itemize}

While stopwords may be related to how \commenters coordinate
their style
with the \OP \cite{Danescu-Niculescu-Mizil+al:12a,Niederhoffer+Pennebaker:2002a}, content
words can be a good signal of new information or new perspectives.
Thus,
inspired by previous results distinguishing these vocabulary types in studying the effect of phrasing \cite{tan+lee+pang:14},
for each of the four variants above
we try three
different word
sets: stopwords, content words and all words.
The features based on interplay
are all significant to a certain degree. 
Similar patterns occur in \fstcomment and \fullpath:
in number of common words and \OP fraction, persuasive arguments have larger values because they tend to be longer, 
as
will be shown in \secref{sec:pair_comment_features};
in \acomment fraction and Jaccard, which are normalized by \acomment length,
persuasive arguments are more dissimilar from the \originalpost in content words but more similar in stopwords.
Keeping in mind that the pairs we compare are chosen to be similar
to each other, our analysis indicates that, under this constraint, persuasive arguments use a more different wording from the \originalpost in content, while at the same time 
matching them
 more on stopwords.

If we instead 
use truncation to (artificially) control for \acomment length,
persuasive arguments
present lower similarity in all metrics,
suggesting that effects might differ over local parts of the texts.
However, it is consistent that successful arguments are less similar to the \originalpost in content words.
\begin{figure*}[t]
	\begin{subfigure}[t]{0.23\textwidth}
        \addFigure{0.92\textwidth}{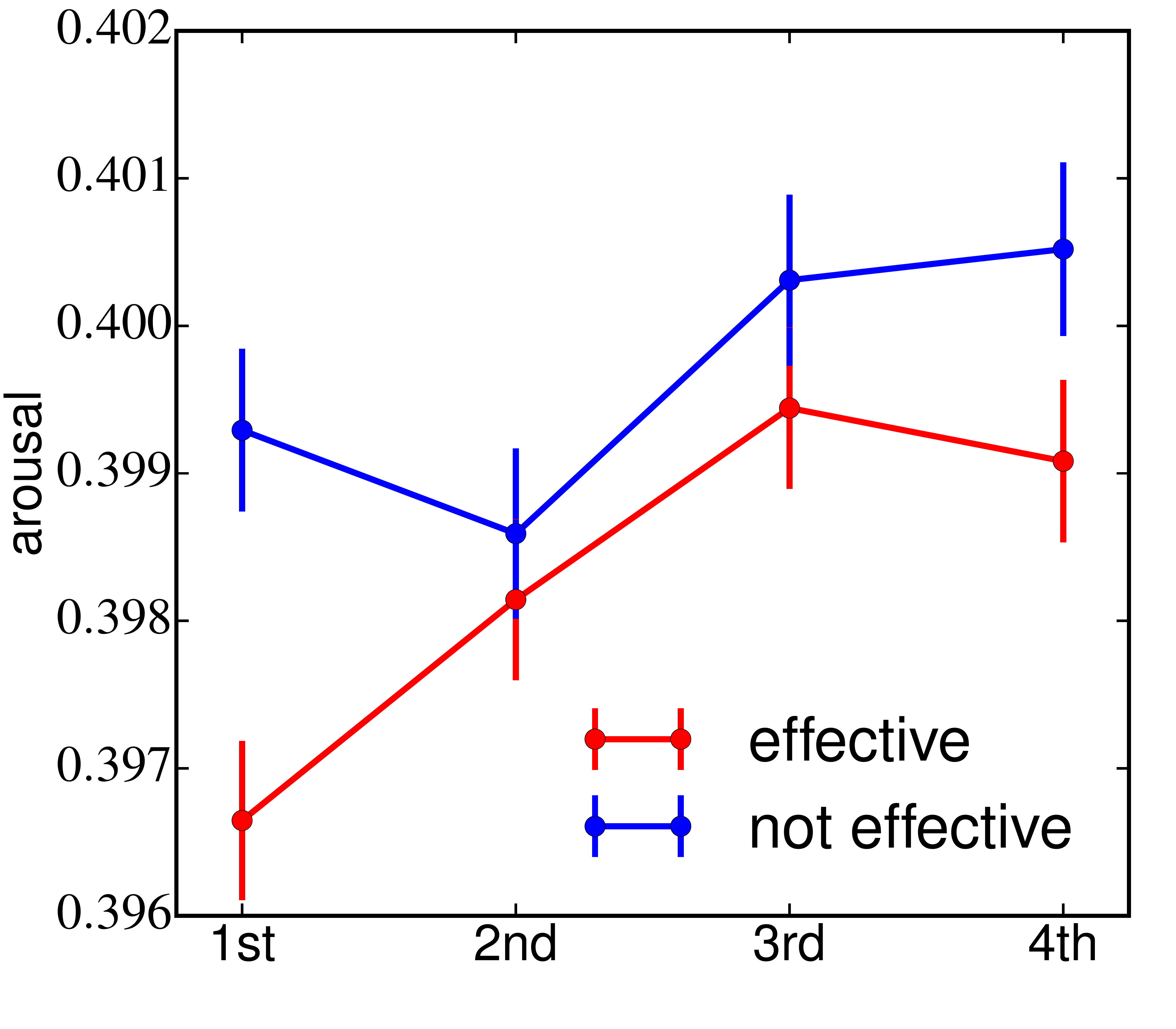}
        \caption{Arousal in \rootcomments.}
        \label{fig:comment_arousal}
    \end{subfigure}
    \hfill
    \begin{subfigure}[t]{0.23\textwidth}
        \addFigure{0.92\textwidth}{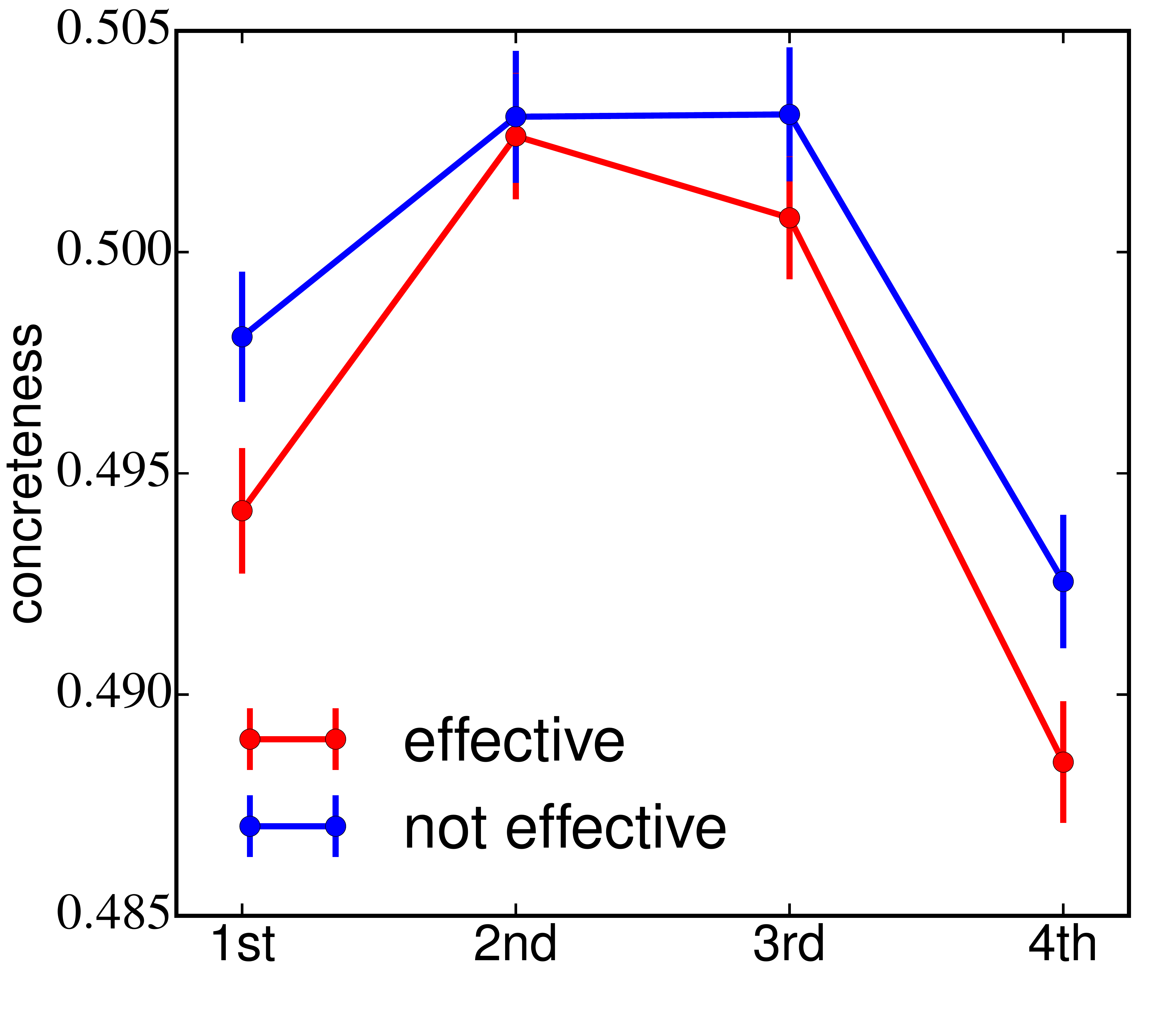}
        \caption{Concreteness in \rootcomments.}
        \label{fig:comment_concreteness}
    \end{subfigure}
    \hfill
    \begin{subfigure}[t]{0.23\textwidth}
        \addFigure{0.92\textwidth}{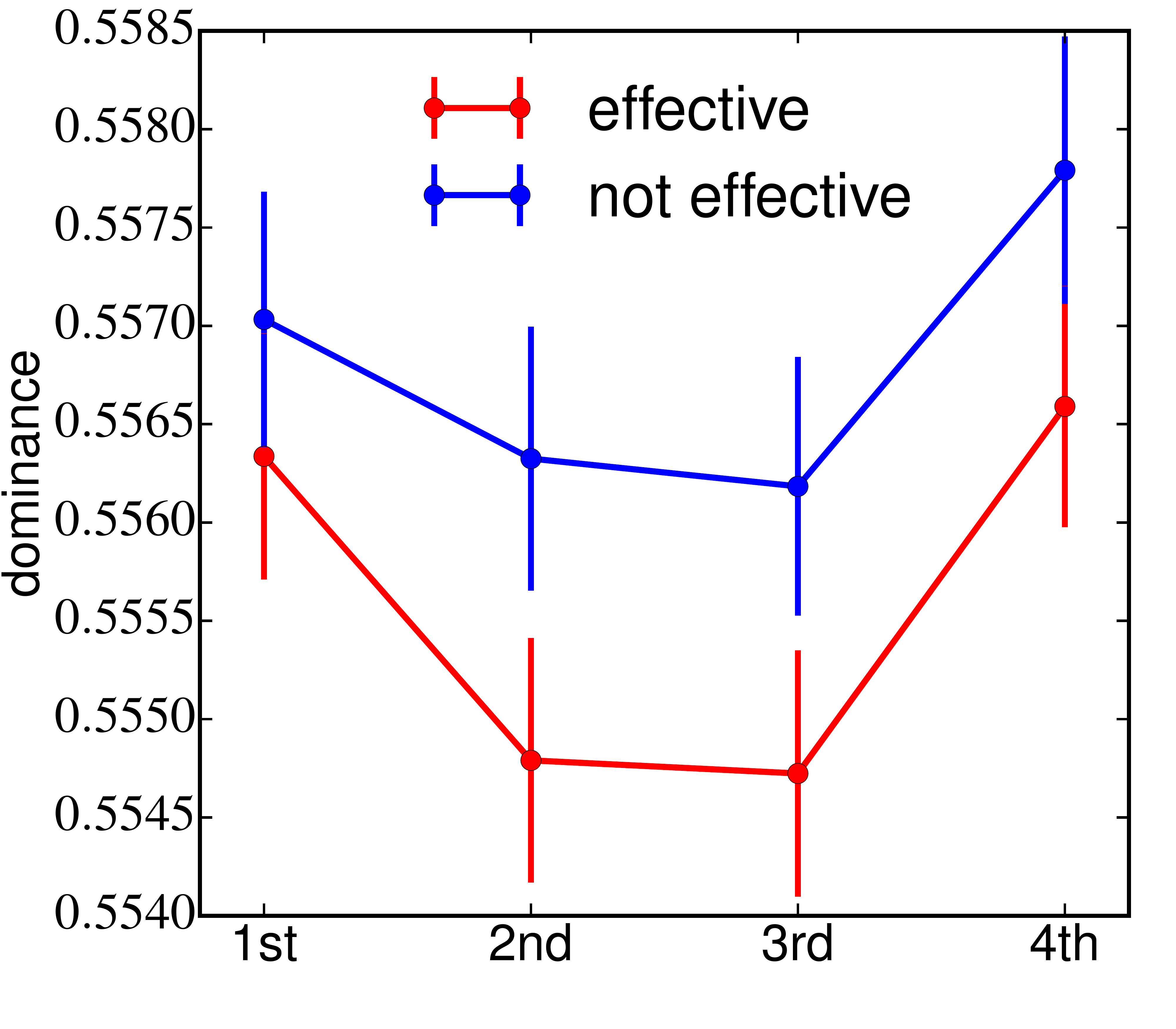}
        \caption{Dominance in \rootcomments.}
        \label{fig:comment_dominance}
    \end{subfigure}
    \hfill
    \begin{subfigure}[t]{0.23\textwidth}
        \addFigure{0.92\textwidth}{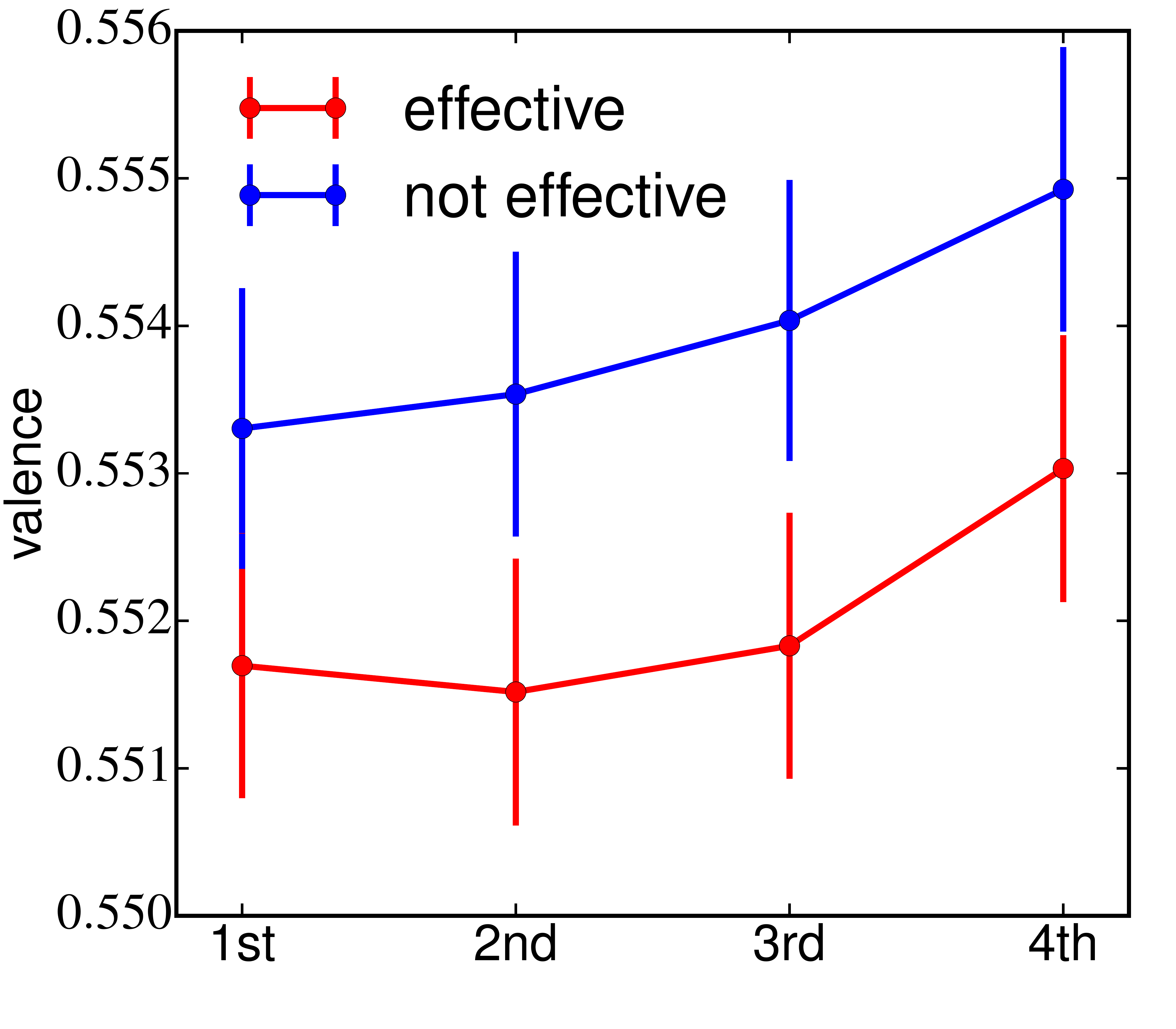}
        \caption{Valence in \rootcomments.}
        \label{fig:comment_valence}
    \end{subfigure}

    \begin{subfigure}[t]{0.23\textwidth}
        \addFigure{0.92\textwidth}{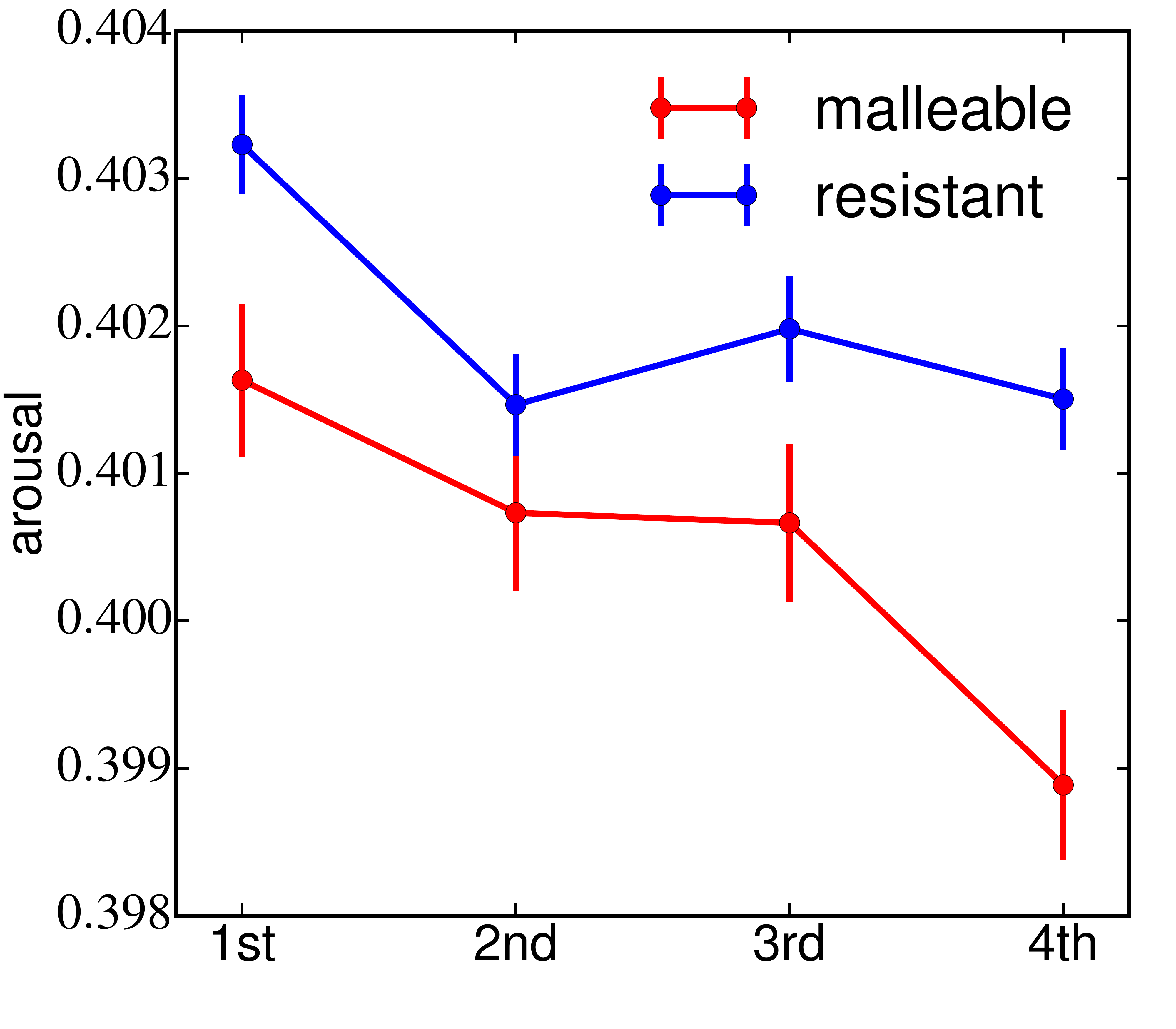}
        \caption{Arousal in {\originalpost}s.}
        \label{fig:op_arousal}
    \end{subfigure}
    \hfill
    \begin{subfigure}[t]{0.23\textwidth}
        \addFigure{0.92\textwidth}{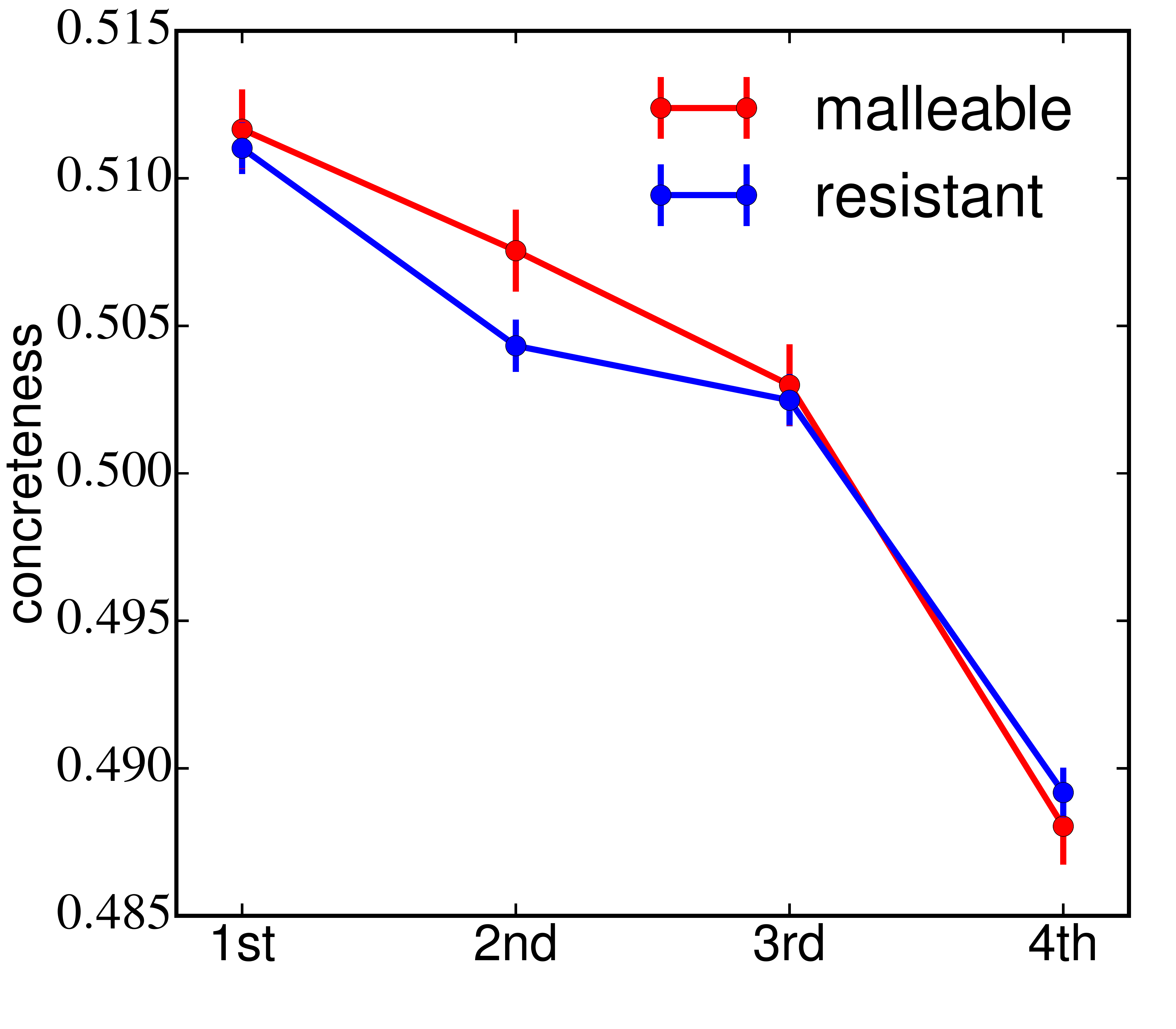}
        \caption{Concreteness in {\originalpost}s.}
        \label{fig:op_concreteness}
    \end{subfigure}
    \hfill
    \begin{subfigure}[t]{0.23\textwidth}
        \addFigure{0.92\textwidth}{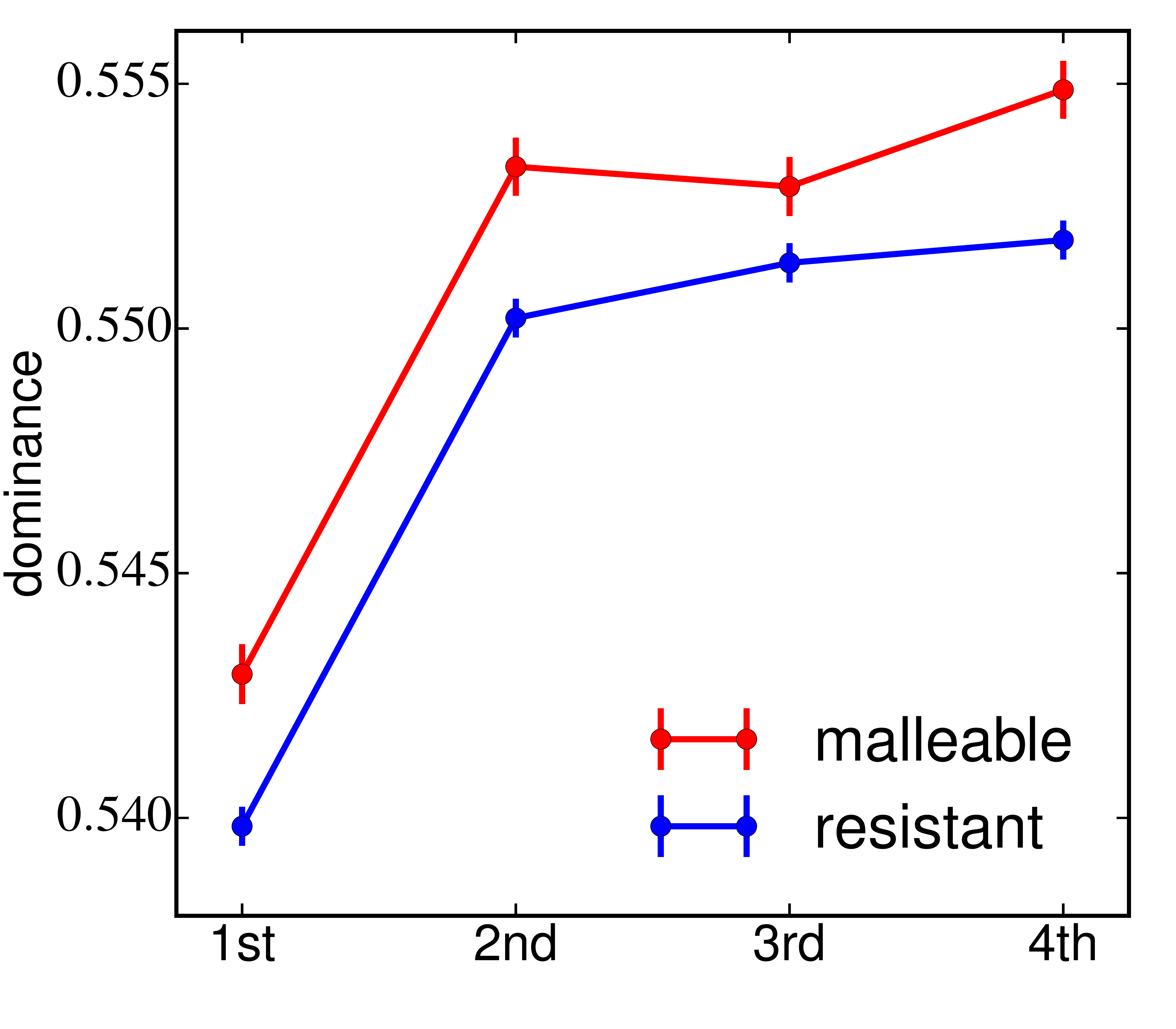}
        \caption{Dominance in {\originalpost}s.}
        \label{fig:op_dominance}
    \end{subfigure}
    \hfill
    \begin{subfigure}[t]{0.23\textwidth}
        \addFigure{0.92\textwidth}{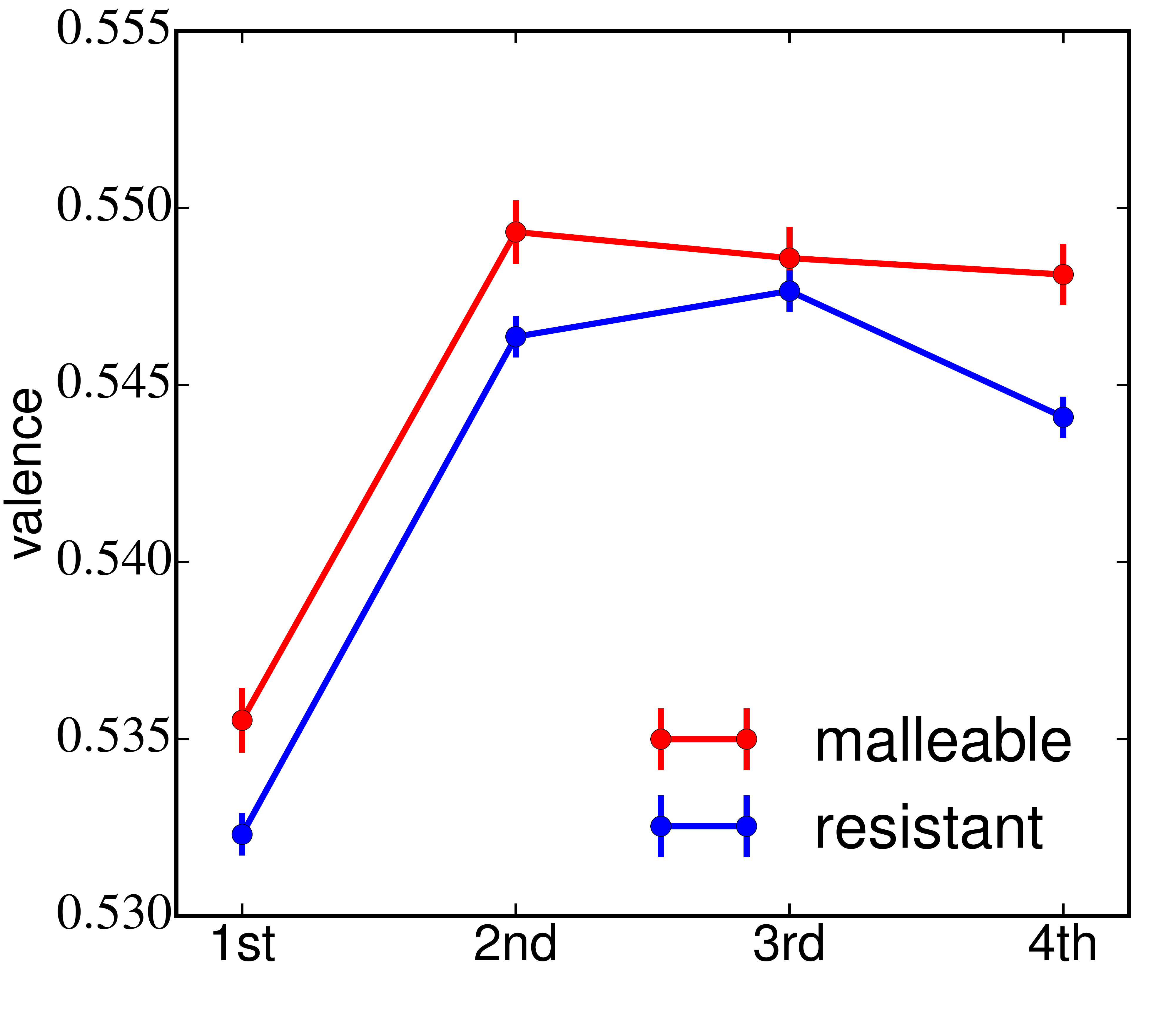}
        \caption{Valence in {\originalpost}s.}
        \label{fig:op_valence}
    \end{subfigure}
    \caption{Style features in different quarters. The first row shows how arousal, concreteness, dominance and valence change in different quarters of the \rootcomment, while the second row shows the
    same features in 
    the {\originalpost}s. 
    The descending concreteness trend suggests that opinions tend to be expressed in a particular-to-general way; \comments notably differ by having both the opening and the closing be abstract, with a concrete middle. These differences are indicative of the functions that the two forms of utterances serve: a \cmv rule is that {\originalpost}s should not be ``like a persuasive essay''.
    Error bars represent standard errors.\label{fig:style_quarter}}
\end{figure*}

\begin{table}[t!]
\centering
\caption{%
Argument-only
features
that pass
a Bonferroni-corrected significance test.
Features are sorted
within each group
by average p-value
over
the
two
tasks.
Due to our simple truncation based on words, some features,
such as those based on complete sentences,
 cannot be
extracted
in \fsttruncated;
these are indicated by
a dash.
We remind the reader of the \fsttruncated disclaimer from \secref{sec:comment_pred}.
\label{tb:pair_comment_test}}
\small
\begin{tabular}{p{1.5in}ll}
\toprule
Feature name & \fstcomment & \fullpath \\
\midrule
\#words & \makebox[17pt][l]{$\uparrow\uparrow\uparrow\uparrow$}  & $\uparrow\uparrow\uparrow\uparrow$ \\

\bigsep
\multicolumn{3}{l}{\bfseries Word category--based features}\\
\smallsep
\#definite articles & \makebox[17pt][l]{$\uparrow\uparrow\uparrow\uparrow$}  & $\uparrow\uparrow\uparrow\uparrow$ \\
\#indefinite articles & \makebox[17pt][l]{$\uparrow\uparrow\uparrow\uparrow$}  & $\uparrow\uparrow\uparrow\uparrow$ \\
\#positive words & \makebox[17pt][l]{$\uparrow\uparrow\uparrow\uparrow$} \textcolor{gray}{($T^R$)} & $\uparrow\uparrow\uparrow\uparrow$ \\
\#2\textsuperscript{nd} person pronoun & \makebox[17pt][l]{$\uparrow\uparrow\uparrow\uparrow$}  & $\uparrow\uparrow\uparrow\uparrow$ \\
\#links & \makebox[17pt][l]{$\uparrow\uparrow\uparrow\uparrow$} \textcolor{gray}{($T$)} & $\uparrow\uparrow\uparrow\uparrow$ \\
\#negative words & \makebox[17pt][l]{$\uparrow\uparrow\uparrow\uparrow$}  & $\uparrow\uparrow\uparrow\uparrow$ \\
\#hedges & \makebox[17pt][l]{$\uparrow\uparrow\uparrow\uparrow$}  & $\uparrow\uparrow\uparrow\uparrow$ \\
\#1\textsuperscript{st} person pronouns & \makebox[17pt][l]{$\uparrow\uparrow\uparrow\uparrow$}  & $\uparrow\uparrow\uparrow\uparrow$ \\
\#1\textsuperscript{st} person plural pronoun & \makebox[17pt][l]{$\uparrow\uparrow\uparrow\uparrow$}  & $\uparrow\uparrow\uparrow\uparrow$ \\
\#\texttt{.com} links & \makebox[17pt][l]{$\uparrow\uparrow\uparrow\uparrow$} \textcolor{gray}{($T$)} & $\uparrow\uparrow\uparrow\uparrow$ \\
frac. links & \makebox[17pt][l]{$\uparrow\uparrow\uparrow\uparrow$} \textcolor{gray}{($T$)} & $\uparrow\uparrow\uparrow\uparrow$ \\
frac. \texttt{.com} links & \makebox[17pt][l]{$\uparrow\uparrow\uparrow\uparrow$} \textcolor{gray}{($T$)} & $\uparrow\uparrow\uparrow\uparrow$ \\
\#examples & \makebox[17pt][l]{$\uparrow$}  & $\uparrow\uparrow\uparrow\uparrow$ \\
frac. definite articles & \makebox[17pt][l]{$\uparrow$} \textcolor{gray}{($T$)} & $\uparrow\uparrow$ \\
\#question marks & \makebox[17pt][l]{$\uparrow$} \textcolor{gray}{---} & $\uparrow\uparrow\uparrow\uparrow$ \\
\#PDF links & \makebox[17pt][l]{$\uparrow$}  & $\uparrow\uparrow\uparrow$ \\
\#\texttt{.edu} links & \makebox[17pt][l]{}  & $\uparrow$ \\
frac. positive words & \makebox[17pt][l]{$\downarrow$}  &  \\
frac. question marks & \makebox[17pt][l]{} \textcolor{gray}{---} & $\downarrow$ \\
\#quotations & \makebox[17pt][l]{}  & $\uparrow\uparrow\uparrow\uparrow$ \\

\bigsep
\multicolumn{3}{l}{\bfseries Word score--based features}\\
\smallsep
arousal & \makebox[17pt][l]{$\downarrow$} \textcolor{gray}{($T$)} & $\downarrow\downarrow\downarrow$ \\
valence & \makebox[17pt][l]{$\downarrow$}  &  \\

\bigsep
\multicolumn{3}{l}{\bfseries Entire argument features}\\
\smallsep
word entropy & \makebox[17pt][l]{$\uparrow\uparrow\uparrow\uparrow$}  & $\uparrow\uparrow\uparrow\uparrow$ \\
\#sentences & \makebox[17pt][l]{$\uparrow\uparrow\uparrow\uparrow$} \textcolor{gray}{---} & $\uparrow\uparrow\uparrow\uparrow$ \\
type-token ratio & \makebox[17pt][l]{$\downarrow\downarrow\downarrow\downarrow$} \textcolor{gray}{($T^R$)} & $\downarrow\downarrow\downarrow\downarrow$ \\
\#paragraphs & \makebox[17pt][l]{$\uparrow\uparrow\uparrow\uparrow$} \textcolor{gray}{---} & $\uparrow\uparrow\uparrow\uparrow$ \\
Flesch-Kincaid grade levels & \makebox[17pt][l]{} \textcolor{gray}{---} & $\downarrow\downarrow\downarrow$ \\

\bigsep
\multicolumn{3}{l}{\bfseries Markdown formatting}\\
\smallsep
\#italics & \makebox[17pt][l]{$\uparrow\uparrow\uparrow\uparrow$} \textcolor{gray}{---} & $\uparrow\uparrow\uparrow\uparrow$ \\
bullet list & \makebox[17pt][l]{$\uparrow\uparrow\uparrow\uparrow$} \textcolor{gray}{---} & $\uparrow\uparrow\uparrow\uparrow$ \\
\#bolds & \makebox[17pt][l]{$\uparrow\uparrow$} \textcolor{gray}{---} & $\uparrow\uparrow\uparrow\uparrow$ \\
numbered words & \makebox[17pt][l]{$\uparrow$}  & $\uparrow\uparrow\uparrow\uparrow$ \\
frac. italics & \makebox[17pt][l]{$\uparrow$} \textcolor{gray}{---} & $\uparrow$ \\

\bottomrule
\end{tabular}
\end{table}

\subsubsection{Argument-only features: \tableref{tb:pair_comment_test}}
\label{sec:pair_comment_features}

We 
now
describe
cues
that can be extracted 
solely
from the \comments.
These features
attempt to capture linguistic
style
and its connections to persuasion success.

\para{Number of words.} A straightforward but powerful feature is the number of words.
In both \fstcomment and \fullpath,
a larger number of words is strongly correlated with success.
This is not surprising: longer
\comments
can
be more explicit \cite{o1997standpoint,o1998justification}
and
convey more information.
But
na\"ively making a communication longer does not automatically make it more
convincing
(indeed, sometimes, more succinct phrasing carries more
punch); our more advanced features
attempt to capture the 
subtler aspects of length.
\para{Word category--based features.}
As suggested by existing psychology
theories and our intuitions, the frequency of
certain
types of
words
may be
associated with
persuasion success.
We consider 
a wide range of categories (see \secref{sec:appendix} for details),
where for each, we measure
the
raw
number of word occurrences
and the length-normalized version.
\para{Word score--based features.}  %
Beyond word categories, we employ four
scalar
word-level attributes \cite{brysbaert2014concreteness,warriner2013norms}:
\begin{itemize}
\item Arousal captures the intensity of an emotion, and ranges from ``calm''
words ({\em librarian, dull}) to words that excite, like {\em terrorism} and
{\em erection}.
\item Concreteness reflects the degree to which a word denotes something
perceptible, as opposed to abstract words which can denote ideas and concepts,
e.g., {\em hamburger} vs. {\em justice}.
\item Dominance measures the degree of control expressed by a word.  Low-%
dominance words can suggest vulnerability and weakness ({\em dementia,
earthquake}) while high-dominance words
evoke
power and success ({\em
completion, smile}).%
\item Valence is a measure of how pleasant the word's denotation is.  Low-%
valence words include {\em leukemia} and {\em murder}, while {\em sunshine}
and {\em lovable} are high-valence.
\end{itemize}
We
scale
the four measures above to lie
in $[0, 1]$.\footnote{While the resources cover
most common words, out-of-vocabulary misses can occur
often in user-generated content.
We found that all four values can be extrapolated with high accuracy to out-of-vocabulary words by regressing on
dependency-based word embeddings \cite{levy2014dependencybased} (median absolute error of about 0.1).
Generalizing lexical attributes using word embeddings was previously used for
applications such as figurative language detection
\cite{tsvetkov2014metaphor}.} We extend these measures to texts by averaging
over the ratings of all content words.
\tableref{tb:pair_comment_test} shows that it is consistently good to
use calmer language.
Aligned with our
findings in
terms of sentiment words (\secref{sec:appendix}),
persuasive arguments are slightly
less happy.
However, no significant differences
were found for
concreteness and dominance.

\para{Characteristics of the entire argument.}
We measure the number of paragraphs and the number of sentences%
:
persuasive arguments have significantly more 
of both.
To capture the lexical diversity in an argument, we consider
the {\em type-token ratio}
and {\em word entropy}.
Persuasive arguments are more diverse in \fstcomment and \fullpath,
but the {\em type-token ratio} is surprisingly higher in \fsttruncated:
because of correlations with length and argument structure, lexical diversity
is hard to interpret for texts of different lengths.
Finally, we compute Flesch-Kincaid grade level \cite{kincaid1975derivation} to represent readability. Although there is no significant difference in \fstcomment, persuasive arguments
are more complex
in \fullpath.

\para{Formatting.}  Last but not least, discussions on the Internet employ
certain writing conventions enabled by the
user
interface.
Since \Reddit comments use Markdown\footnote{
\small{\url{https://daringfireball.net/projects/markdown/}}
} for formatting,
we can recover the usage of bold, italic,
bullet lists, numbered lists and links formatting.\footnote{We also consider
numbered words
({\em first}, {\em second}, {\em third}, etc.)
as the textual version of numbered
lists.}  While these features are not applicable in face-to-face arguments,
more and more communication takes place online, making them highly relevant.
Using absolute number, most
of them
are significant except numbered lists.
When it comes to normalized counts, though,
only
italicizing
exhibits significance.

\begin{figure*}[t]
\centering
\begin{tabular}{p{0.32\textwidth}p{0.68\textwidth}}
\begin{minipage}[t]{0.3\textwidth}
    \addFigure{\textwidth}{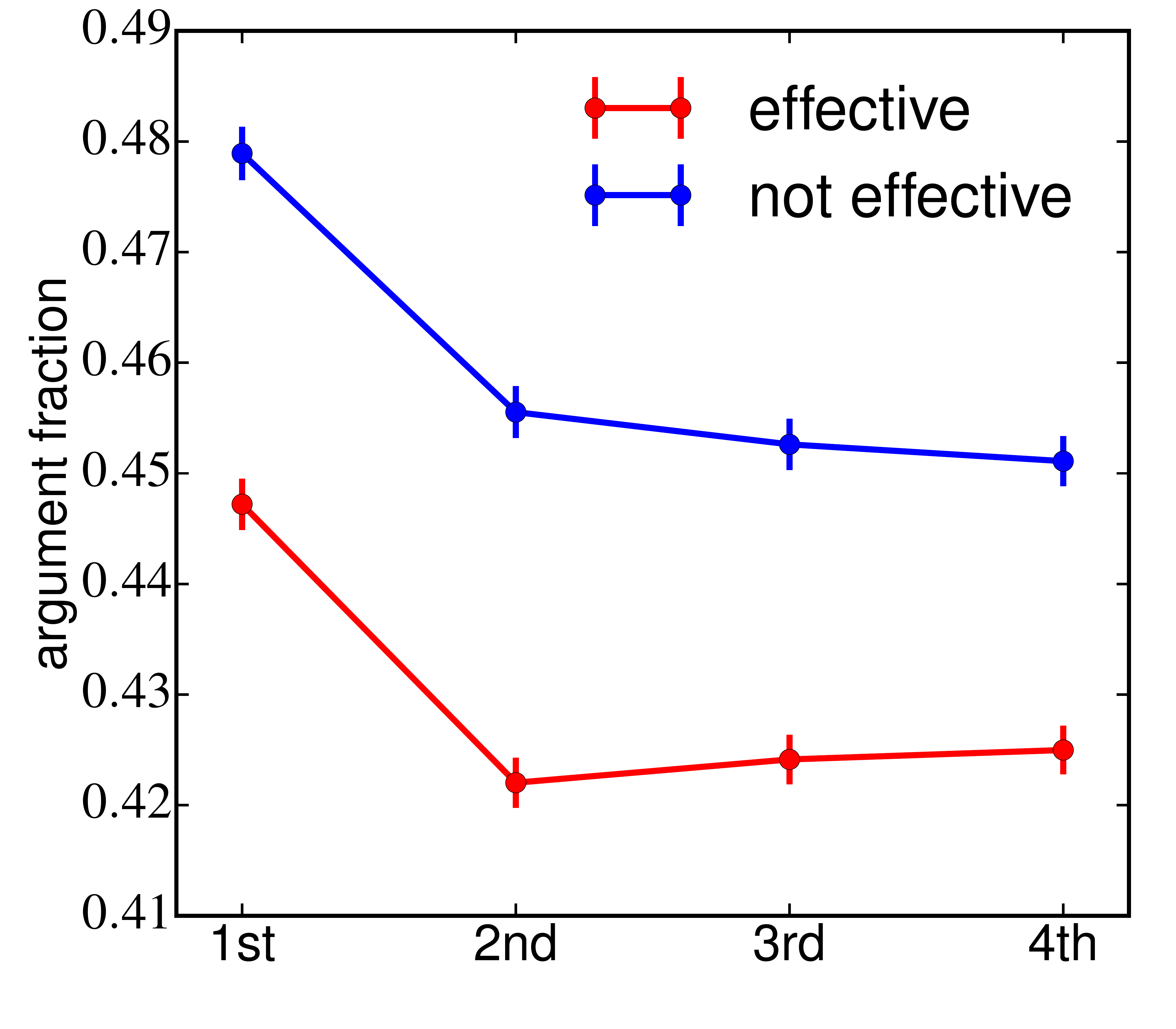}
    \caption{Similarity between each quarter of an argument and the entire
    {\originalpost}.}
    \label{fig:simi_quarter}
\end{minipage}
&
\begin{minipage}[t]{0.68\textwidth}
    \centering
    \addFigure{0.95\textwidth}{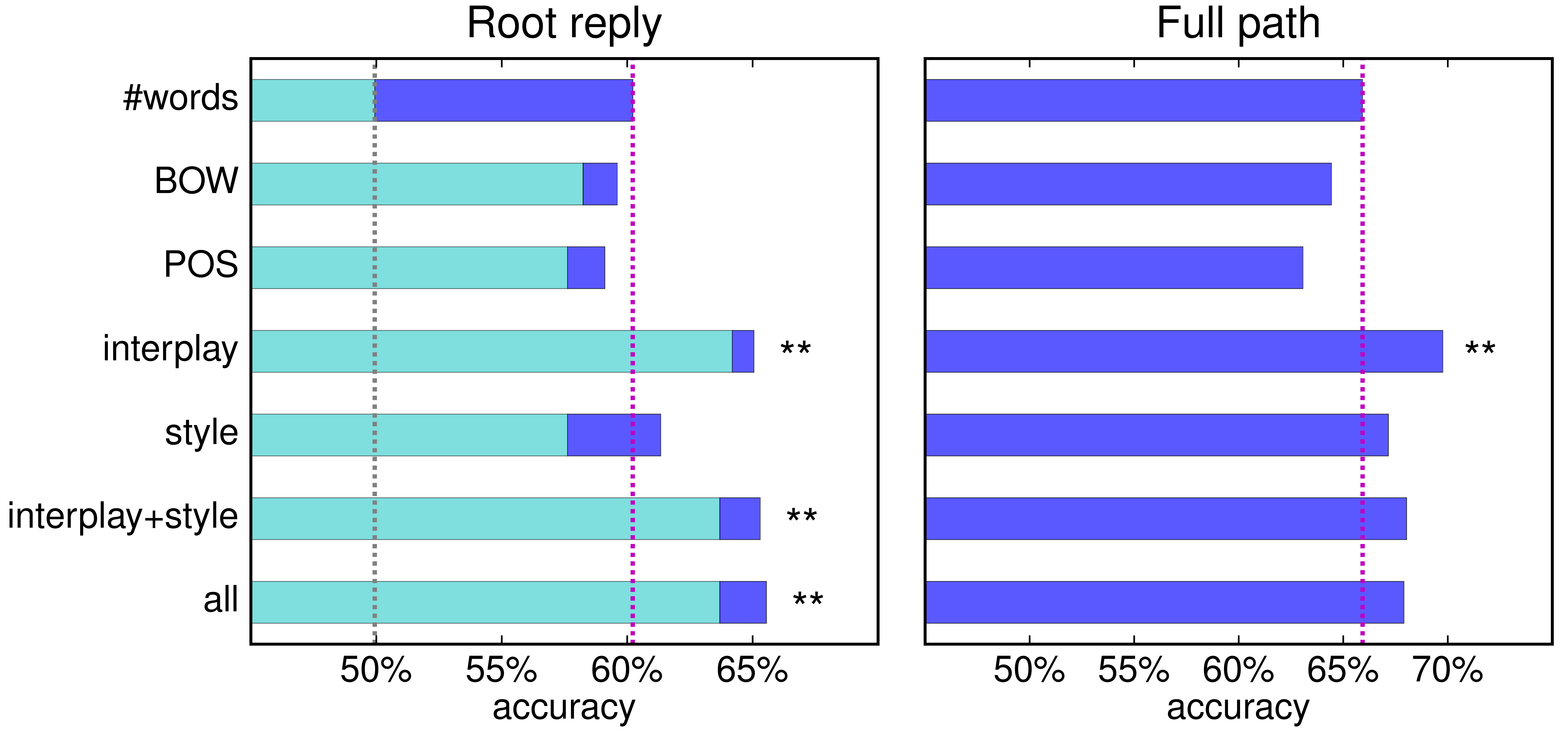}
    \caption{
    {\bf Prediction results.}
    The cyan fraction in the left figure shows the performance in \fsttruncated, and the purple bar shows the performance in \fstcomment.
    The magenta line shows the performance of {\em \#words} in \fstcomment, while the gray line shows the performance of {\em \#words} in \fsttruncated, which is the same as random guessing.
    The
    figure on the right
    gives the performance in \fullpath (the magenta line gives the performance of {\em \#words}).
    The number of stars indicate the significance level compared to the
    {\em \#words} baseline
    according to McNemar's test.
    (*: $p < 0.05$, **: $p < 0.01$, ***: $p < 0.001$.)}
    \label{fig:pair_prediction}
\end{minipage}
\end{tabular}
\end{figure*}

\subsubsection{They hold no quarter, they ask no quarter}
\label{sec:pair_comment_quarters}

Understanding
how a line of argument might evolve is
another
interesting research problem.
We investigate by
quartering each
argument
and measuring
certain feature values in each quarter, allowing for finer-grained
insight into argument structure.

\para{Word score--based features in quarters.}  %
(\figref{fig:style_quarter})
With the exception of arousal,
effective arguments and ineffective arguments present similar patterns:
the
middle is more concrete and less dominant than the beginning and end,
while valence rises slightly over the course of an argument. We also see
interesting differences in psycholinguistic patterns between \plural{\originalpost}
and \comments. (We defer detailed discussion to \secref{sec:op_pred}.)  In terms
of arousal, however, successful arguments begin by using calmer words.

\para{Interplay with
the
\originalpost.} (\figref{fig:simi_quarter}) To capture partial overlap and possible divergence
from the OP's view, we divide both the
\originalpost
and the \rootedpathunit into quarters, and
measure similarity metrics between all subdivisions (including the full
unit).\footnote{In prediction, we also take the maximum and minimum of these
quarter-wise measures as an order-independent way to summarize fragment
similarity.}
Since the \acomment fraction in content words is the most significant interplay
feature, in \figref{fig:simi_quarter} we only show the fraction of common content words in different
quarters of \comments vs. the original post. Both
effective and ineffective arguments
start off
more similar with
the \originalpost;
effective arguments remain less similar overall.

\subsection{Prediction results}
\label{sec:pred_results}

We train logistic regression models with $\ell_1$ regularization on the
training set and choose parameters using five cross-validation folds, ensuring
that all pairs of arguments that share the same OP are in the same
fold.\footnote{We also tried $\ell_2$ regularization, random forests and
gradient boosting classifiers and found no improvement beyond the cross-%
validation standard error.
} All features are standardized to
unit variance, and missing values are imputed using the training fold sample
mean. We evaluate using
pairwise accuracy
in the {\em heldout dataset},
where we restricted ourselves to \emph{a single experimental run} (after holding
our
collective breath)
to further reduce the 
risk
of overfitting.
The results are, in fact, in line
with what we 
describe
 in the training-data analysis here.

\para{Feature sets.}
As shown in \secref{sec:pair_features},
the number of words
is very predictive,
providing a strong baseline to compare against.
Bag-of-words
features (\BOW)
usually
provide
a strong benchmark
for text classification tasks. We restrict the size of the vocabulary by
removing rare words that occurred no more than 5 times in training and
$\ell_2$-normalize term frequency vectors.  %
Since part-of-speech tags may also capture properties of the argument, we also
use normalized term frequency vectors by treating part-of-speech tags as words
(\POS). 
Features in \secref{sec:pair_simi_features} are referred to
as {\em interplay};
features in \secref{sec:pair_comment_features} constitute 
the 
feature
set 
{\em style}.
Finally, we use a combination 
of 
style and interplay, as well
as a combination that includes all the above features ({\em all}). Note that
style and interplay
are dense and very low-dimensional
compared to \BOW.

\para{Interplay with
the
OP plays an essential
role.}
(\figref{fig:pair_prediction}%
)
{\em \#words}
is indeed a very strong baseline that achieves an accuracy of 60\% in \fstcomment and 
66\%
in \fullpath.
As a sanity check, in \fsttruncated, it indeed gets only 50\%. In comparison,
\BOW
achieves
similar performance as
{\em \#words},
while \POS gives even 
worse
 performance.
However, interplay features lead to a 5\% 
absolute
improvement
over the {\em \#words} baseline
in \fstcomment and \fullpath, and a 14\% 
absolute
improvement in \fsttruncated.
In fact, the performance of interplay is already close to using the
combination of interplay and style and using all features. 
In \fsttruncated,
although
the performance of style features drops
significantly, %
interplay achieves very similar performance
as in \fstcomment,
demonstrating
the robustness of the interplay features.

\section{``Resistance'' to persuasion}
\label{sec:op_pred}

Although it is a good-faith step for
a person
to post 
on
\cmv,
some beliefs
in the dataset are
still ``resistant'' to changes,
possibly  %
depending on how strongly the \OP
holds
them and how the \OP acquired and maintained them
\cite{Pomerantz:JournalOfPersonalityAndSocialPsychology:1995,tormala2002doesn,zuwerink1996attitude}.
Since \cmv members must state their opinion and reasons for it in their own words, we can investigate differences between how \resist and \susc views are expressed.
In this section, we seek linguistic and style patterns characterizing 
{\originalpost}s in order to better understand the mechanisms behind attitude
resistance and expression, and to give
potential
\commenters a sense of which views
may be
\resist
before they engage.
However, recognizing the ``\susc'' cases is not an easy task: in a pilot study, human annotators perform at chance level (50\% on a paired task to distinguish which of two {\originalpost}s is \susc).
In light of our observation that persuasion is unsuccessful in 70\% of the cases
from
\secref{sec:exploration}, we
set up an imbalanced prediction task.
We focus on cases where at least 10 \commenters attempt
counterarguments, and
where
the \OP replied at least once,\footnote{Although in preprocessing we 
replaced all explicit edits, we 
also
remove all posts
containing
the word {\em ``changed''}, to avoid 
including post-hoc signals of view change.
}
alleviating the 
concern that an opinion appears resistant simply because there was little effort towards changing it.
This brings us 10,743 {\originalpost}s in the training data and 1,529 {\originalpost}s in the heldout data.
We 
then analyze systematic expression patterns that characterize \susc beliefs
and
that
signal open-mindedness.

\subsection{Stylistic features for open-mindedness}
We employ the same set of features from \secref{sec:pair_comment_features} to capture the characteristics of {\originalpost}s.
Among them, only a handful are
significantly predictive of malleability, as shown in \tableref{tb:op_test}.

\para{Personal pronouns and self-affirmation.}
First person pronouns are strong indicators of malleability, but first person plural pronouns 
correlate with
resistance.  In psychology, self-affirmation has been found to indicate open-mindedness and make beliefs more likely to yield \cite{cohen2000beliefs,correll2004affirmed}. Our result aligns with these findings: individualizing one's relationship with a belief using first person pronouns affirms the self, while first person plurals can indicate a
diluted sense of group responsibility
for the view.
Note that it was also found in other work that openness is negatively correlated with first person singular pronouns \cite{Pennebaker:JPersSocPsychol:1999}.

\begin{table}
\centering
\caption{Opinion malleability task: statistically significant features after Bonferroni correction.\label{tb:op_test}}
\small
\begin{tabular}{ll}
\toprule
Feature name & More malleable? \\
\midrule  %
\#1\textsuperscript{st} person pronouns & $\uparrow\uparrow\uparrow\uparrow$ \\
frac. 1\textsuperscript{st} person pronoun & $\uparrow\uparrow\uparrow\uparrow$ \\
dominance & $\uparrow\uparrow\uparrow\uparrow$ \\
frac. 1\textsuperscript{st} person plural pronoun & $\downarrow\downarrow\downarrow$ \\
\#paragraphs & $\uparrow\uparrow$ \\
\#1\textsuperscript{st} person plural pronoun & $\downarrow\downarrow$ \\
\#bolds & $\uparrow$ \\
arousal & $\downarrow$ \\
valence & $\uparrow$ \\
bullet list & $\uparrow$ \\

\bottomrule
\end{tabular}
\vspace{-0.1in}
\end{table}

\para{Formatting.}  The use of more paragraphs, bold formatting, and
bulleted lists are all higher when a \susc view is expressed.  Taking more
time and presenting the reasons behind an opinion in a more elaborated form
can indicate
more engagement.
\para{Word score--based features.}
Dominance is the most predictive of malleability: the average amount
of control expressed
through the words used is higher when
describing a \susc view than a \resist one.  The same holds for happiness
(captured by valence).  In terms of arousal, \susc opinions are expressed
significantly more serenely, ending on a particularly
calm note in the
final quarter, while stubborn opinions are expressed with relatively more
excitement. 

\subsection{Prediction performance}

\begin{figure}[t]
    \centering
    \addFigure{0.3\textwidth}{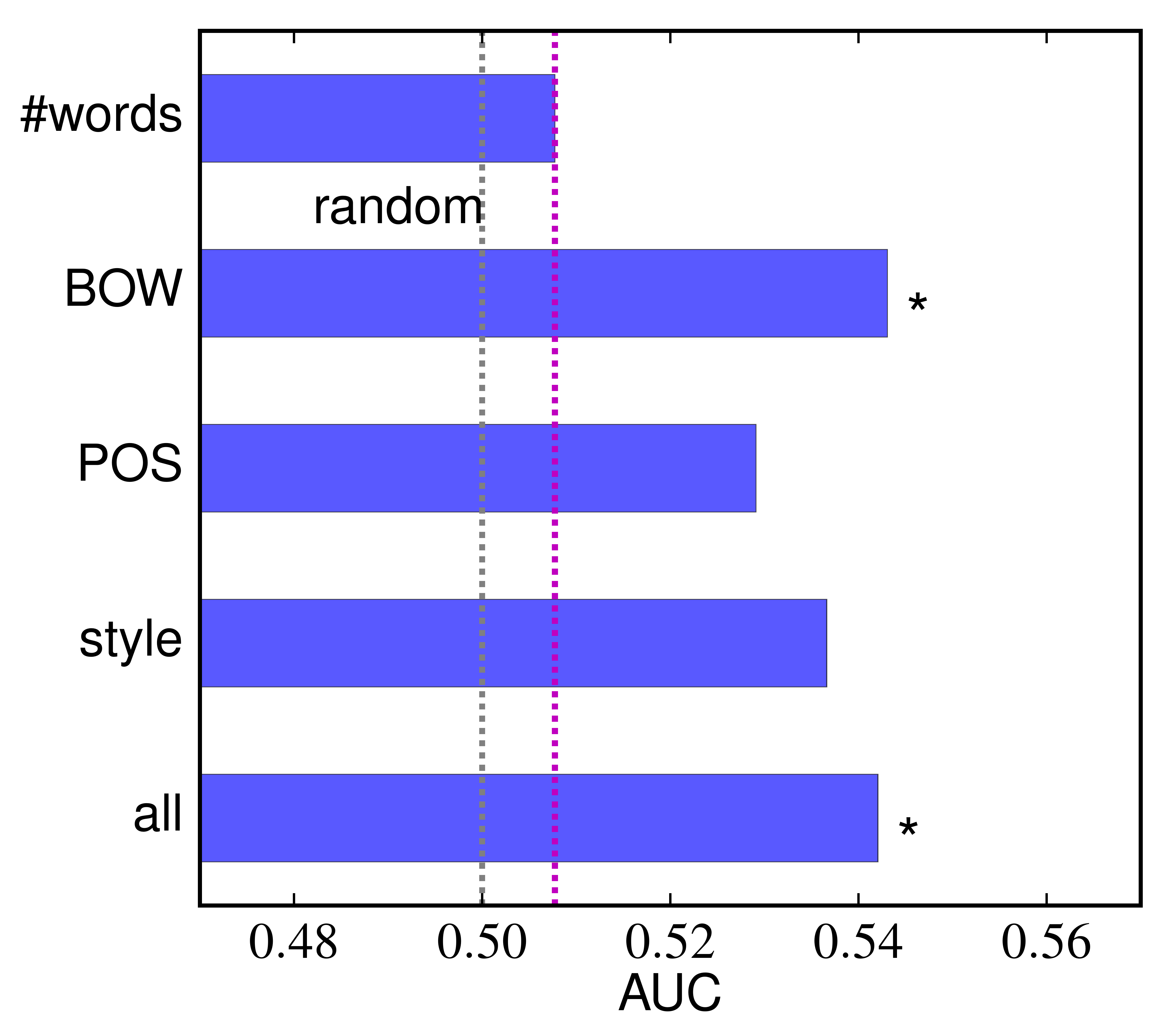}
    \caption{{\bf Opinion malleability prediction performance}:
    AUC on the heldout dataset.
    The purple line shows the performance of
    {\em \#words}, while the gray line gives the performance of random guessing.
    The {\em BOW} and {\em all} feature sets perform significantly better than
    the 
    {\em \#words}
    baseline, according to one-sided
    paired permutation tests. 
    \BOW, \POS, {\em style} and {\em all} outperform random guessing using bootstrapped tests.
    (*: $p < 0.05$, **: $p < 0.01$, ***: $p < 0.001$.)
    \label{fig:op_pred}}
\end{figure}

We use 
weighted 
logistic regression and choose the amount and type of regularization
($\ell_1$ or $\ell_2$) by grid search over 5 cross-validation folds.  Since
this is an imbalanced 
task, we evaluate the prediction results using the
area under the ROC curve (AUC) score.  
As in \secref{sec:comment_pred}, we use the number of words as our baseline.
In addition to the above features that characterizes language style ({\em style}),
we use bag-of-words (\BOW), part-of-speech tags (\POS)
and a full feature set ({\em all}).
The holdout performance is shown in
\figref{fig:op_pred}.

The 
classifiers trained on
bag of words features significantly outperforms the {\em \#words} baseline.
Among words with largest coefficients,
\resist views tend to be expressed using more decisive words such as
{\em anyone},
{\em certain},
{\em ever},
{\em nothing},
and {\em wrong},
while {\em help} and {\em please} are \susc words.
The \POS classifier significantly outperforms random guessing,
but not the
baseline.
Nevertheless, it 
yields an
interesting insight:
comparative adjectives and adverbs are signs of malleability, while superlative adjectives suggest stubbornness.
The full feature set ({\em all}) also 
significantly outperform the 
{\em \#words}
baseline. 
The overall low
scores suggest
that this is indeed a challenging task
for both humans and machines.

\section{%
Further discussion}
\label{sec:discussion}

\begin{figure}[t]
\centering
	\begin{subfigure}[t]{0.23\textwidth}
		\addFigure{\textwidth}{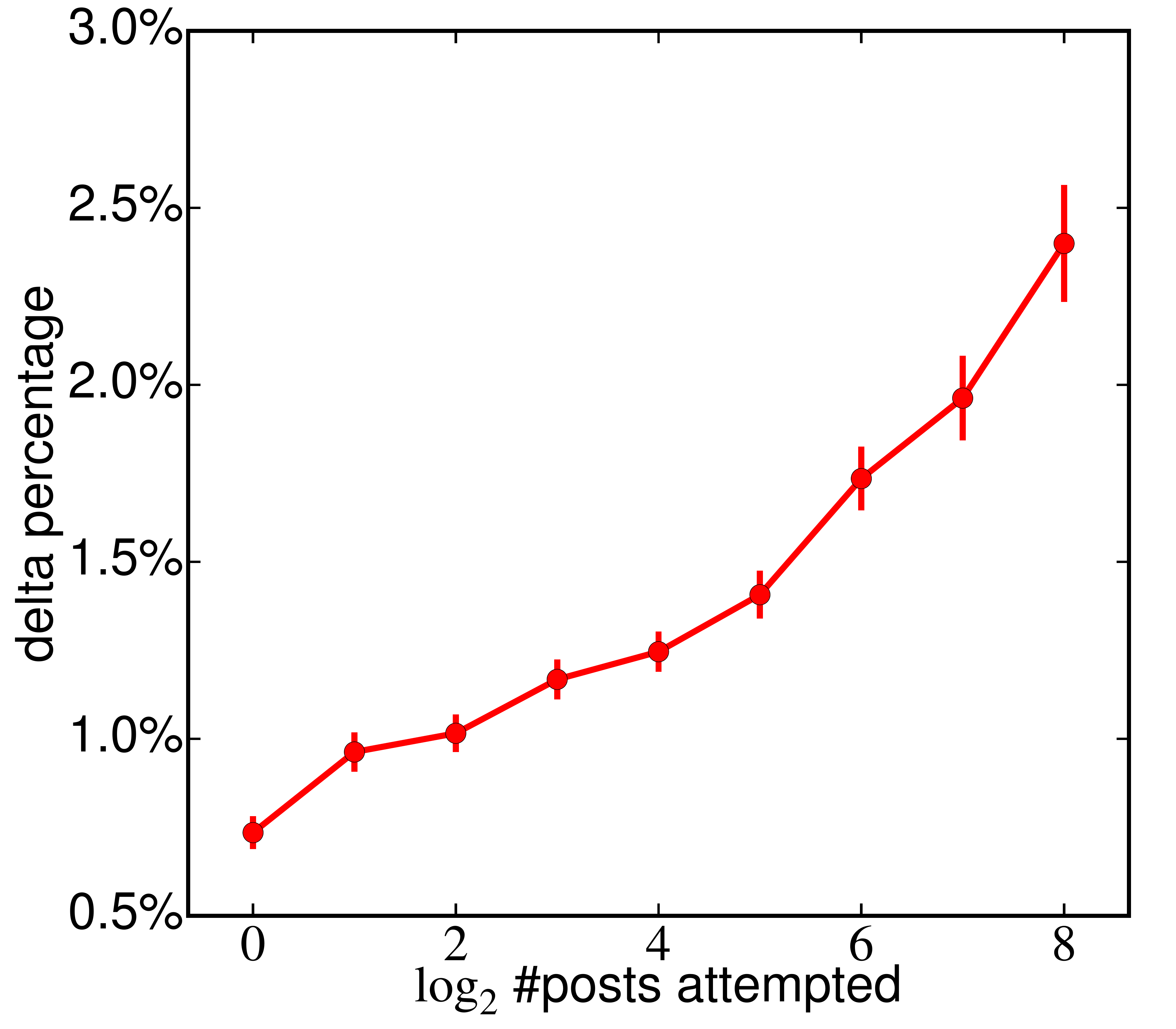}
		\caption{Delta ratio vs. $\log_2$ \#posts.}
		\label{fig:delta_experience}
	\end{subfigure}
	\hfill
	\begin{subfigure}[t]{0.23\textwidth}
		\addFigure{\textwidth}{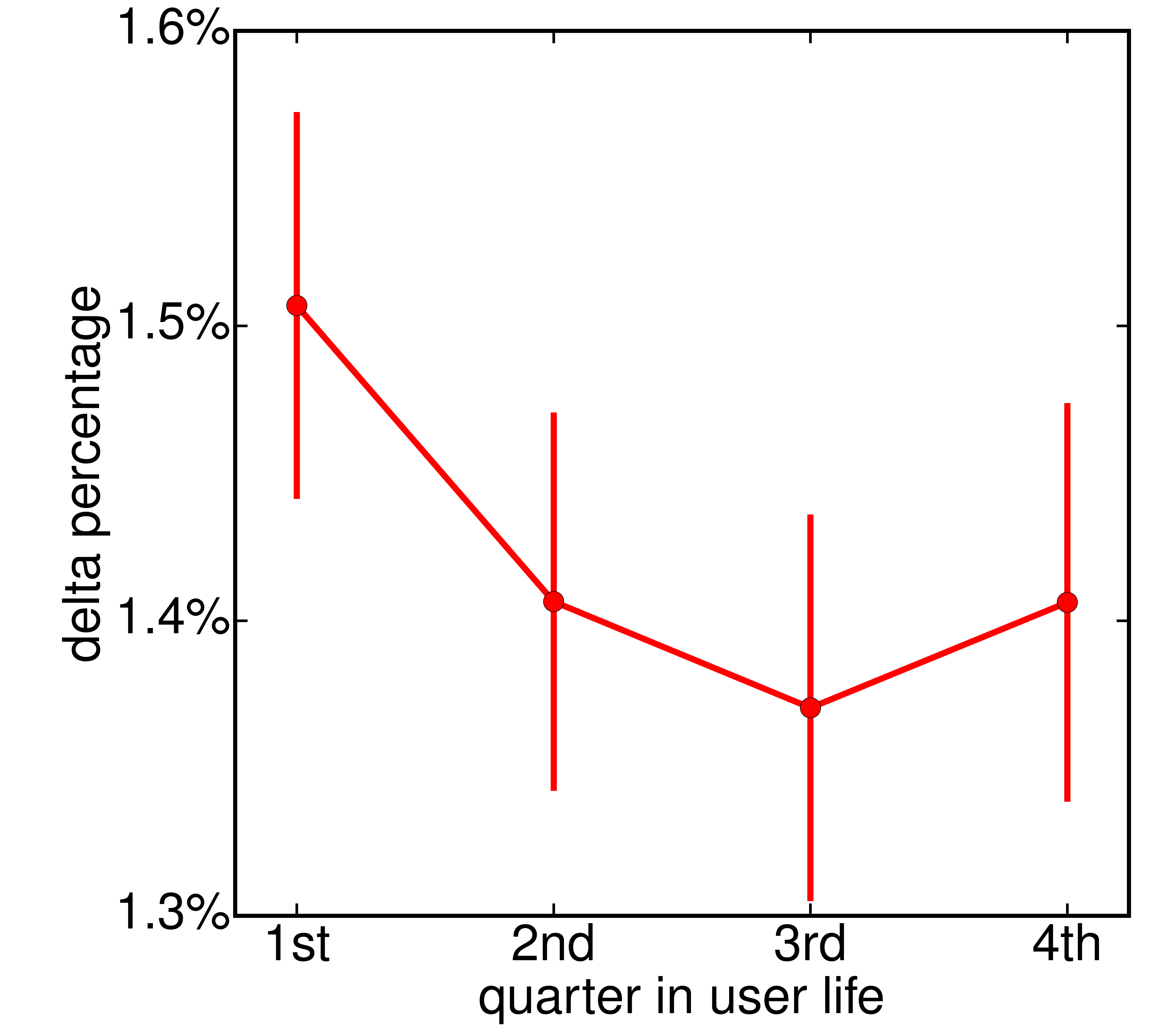}
		\caption{No growth over life.}
		\label{fig:delta_life}
	\end{subfigure}
	\caption{Effect of experience.\label{fig:experience}}
\end{figure}

Here we discuss other observations that
may
open up avenues for further investigation of the complex process of persuasion.
\para{Experience level.}
Beyond the interactions within a \discussiontree, 
\cmv is a community where users can accumulate experience and potentially improve their
persuasion ability.
\figref{fig:delta_experience} 
shows that
a member's
success rate goes up with the number of attempts
made.
This observation can be explained by at least two reasons:
the success rate of frequent \commenters improves over
time, and/or
frequent \commenters are better at persuasion from the beginning.
To disentangle these two possible reasons,
for 
\commenters
who attempted
to change at least 16 views,
we split all the
attempts
into 4 equal chunks sorted by time.
\figref{fig:delta_life} presents how the success rate changes over a \commenter's life,
suggesting
that the success rate of frequent \commenters does not
increase.%
\footnote{%
In terms of the correlation between previous success (lifetime deltas)
and success rate, the result is similar:
beyond 4--5 deltas there is no noticeable increase.
}
It is worth noting that this lack of apparent improvement might be explained by a gradual development of a ``taste'' for {\originalpost}s that are harder to address \cite{McAuley:2013:ACM:2488388.2488466}.
Such community dynamics point to interesting research questions for future work.

\para{Attempts to capture high-level linguistic properties.}
We experimented with a broader set of features in cross validation, which we still deem interesting but did not 
discuss in depth
for space reasons.
One important class are attempts to capture the semantics of original statements and arguments. 
We experimented with using topic models \cite{Blei:2003:LDA:944919.944937} to find topics that are
the
most \susc ({\em topic: food, eat, eating, thing, meat} and {\em topic: read, book, lot, books, women}), and
the
most \resist ({\em topic: government, state, world, country, countries} and {\em topic: sex, women, fat, person, weight}).
However, topic model based features
do not seem to bring predictive power to either of the tasks.
For predicting persuasive arguments,
we attempted to capture interplay with
word embeddings
for text similarity
using both
the centroid distance and
the word mover's distance \cite{kusner2015doc}. 
Both distances
proved predictive by themselves, but
were not able to improve over the features presented in the paper in cross validation.
More generally, better semantic models applicable to online discussions could open up deeper investigations into effective persuasion strategies.
\para{Sequential argument structure.} Another 
promising direction is to examine the structure of arguments via the sequence of discourse connectors.
For instance, we can recover interesting structures such as
``{\em first}$_0$--{\em but}$_1$--{\em because}$_2$''
and
``{\em now}$_1$--{\em then}$_2$--{\em instead}$_3$'',
where the
subscripts
indicate which quarter the discourse connector occurred in.
These features
did not perform well
in our tasks due to low recall, or lack of
argumentative
structure in
the data,
but they 
deserve
 further exploration.

\section{Additional related work}
\label{sec:related}
A few lines of research in natural language processing are related to our work. Argumentation mining focuses on fine-grained analysis
of arguments
and
on discovering the relationships, such as support and premise, between different arguments \cite{Mochales+Moens:11a}.
Studies have also worked on understanding persuasive essays \cite{farra15scoring,persing2015modeling,stab2014identifying} and opinion analysis in terms of agreement and ideology \cite{Thomas+Pang+Lee:06a,somasundaran2010recognizing,hasan-ng:2014:EMNLP2014,sridhar2015joint,rosenthal2015couldn}. Another innovative way of using Internet
data
to study
mass
persuasion is
through
AdWords \cite{guerini2010evaluation}.
\section{Conclusion}
\label{sec:conclusion}

In this work, in order to understand the mechanisms behind persuasion, we
use
a unique dataset from 
\fullcmv.
In addition to examining interaction dynamics, we develop a framework for analyzing persuasive arguments and malleable opinions.
We find that not only are interaction patterns connected to the success of persuasion,
but language is also found to
distinguish persuasive arguments.
Dissimilarity with the wording in which the opinion is expressed turns out to be the most predictive signal among all features.
Although 
members of \cmv are open-minded and willing to change,
we are still able to identify opinions that are
\resist and to characterize them using linguistic
patterns.

There are many possible extensions 
to 
our approach 
for representing arguments.
In particular, it would be interesting to model the framing of different arguments and examine the interplay between framing of the \originalpost and the \comments. For instance, is benefit-cost analysis the only way to convince a utilitarian?

Furthermore, although this novel dataset opens up potential opportunities for future work,
other environments, where people are not as open-minded, can 
exhibit
different kinds of
persuasive interactions;
it
remains an interesting problem how our findings generalize to different contexts.
It is also important to understand
the effects of attitude change on actual behavior \cite{petty1997attitudes}.
Finally, beyond mechanisms
behind
persuasion, it is a vital research problem to understand how community norms encourage such a well-behaved platform so that useful rules,
moderation practices, or even automated tools
can be
deployed in
future community building.

\section{Appendix}
\label{sec:appendix}
In this section
we explain the
features based on word categories.

\begin{itemize}[leftmargin=0.1in,itemsep=0pt]
\item (In)definite articles
(inspired by \cite
{Danescu-Niculescu-Mizil+Cheng+Kleinberg+Lee:12}).
These are highly correlated with
length, so they are both
highly
significant in terms of absolute numbers. However,
in terms of word ratios, definite articles 
(e.g., ``the'' instead of ``a'')
are preferred, which suggests that
specificity is important in persuasive arguments.

\item Positive and negative words.
We use the positive and negative lexicons from LIWC \cite{pennebaker2001linguistic}. In absolute
numbers,
successful arguments 
are more sentiment-laden in both \fstcomment and \fullpath.
When truncating, as well as when taking the frequency ratio,
persuasive opening arguments use {\em fewer} positive words, suggesting
more complex patterns of positive emotion in longer arguments \cite{hullett2005impact,wegener1996effects}.

\item
Arguer-relevant
personal pronouns. We consider
1\textsuperscript{st}
person pronouns
({\em me})
2\textsuperscript{nd}
person pronouns
({\em you})
and
1\textsuperscript{st}
person plural pronouns
({\em us}).
In both \fstcomment and
\fullpath, persuasive arguments use
a significantly larger absolute number of personal pronouns.

\item
Links.  Citing external evidence online is often accomplished
using hyperlinks.
Persuasive arguments use consistently more links, both in absolute 
and in
per-word count.
We make special categories for interesting classes of links: those to {\tt
.com} and {\tt .edu} domains, and those to PDF documents.
Maybe due to
high
recall, {\tt .com} links seem to be most powerful.
Features based on links also tend to be significant even in
the \fsttruncated condition.

\item Hedging.
Hedges indicate uncertainty; 
an example is
``It could be the case''.
Their presence might signal a weaker argument \cite{Durik:JournalOfLanguageAndSocialPsychology:2008}, but alternately, they
may make an argument easier to accept by softening its tone
\cite{lakoff1975hedges}.
We curate a set of hedging cues based on \cite{hanauer2012hedging,hyland1998hedging}.
Hedging is more common in persuasive arguments under
\fstcomment and \fullpath.

\item Examples.
We consider
occurrences of
``for
example'', ``for instance'',
and
``e.g.''.
The absolute number of such example markers is significantly higher in persuasive arguments.

\item Question marks. Questions can be used for clarification or rhetorical
purposes. In terms of absolute number, there are more in \fstcomment and
\fullpath. But
when it comes to ratio, if anything,
it seems better to avoid using
question marks.

\item Quotations. One common practice
in argumentation is to quote the other party's words. However, this does not seem to be a useful strategy
for the
\rootcomment.

\end{itemize}

\newcommand{\finit}[2]{#1.}
\para{Acknowledgments.}   We thank 
\finit{X}{ilun} Chen,
\finit{J}{ack} Hessel,
\finit{S}{ofia} Hu,
\finit{S}{hohini} Kundu,
\finit{N}{ingzi} Li,
\finit{M}{aithra} Raghu,
\finit{N}{athaniel} Rojas,
\finit{A}{lexandra} Schofield,
\finit{T}{ianze} Shi,
and \finit{J}{ustine} Zhang
for participating in our pilot annotation experiments. 
We thank 
\finit{Y}{oav} Artzi,
\finit{J}{ordan} Carpenter,
\finit{P}{rzemyslaw} Grabowicz,
\finit{J}{ack} Hessel,
\finit{Y}{iqing} Hua,
\finit{J}{on} Park,
\finit{M}{ats} Rooth,
\finit{A}{lexandra} Schofield,
\finit{A}{shudeep} Singh,
and the anonymous reviewers 
for helpful comments.
This work was
supported in part by NSF grant IIS-0910664, a Google Research
Grant, a Google Faculty Research Award and a Facebook Fellowship.
\eject\vfill

\bibliographystyle{abbrvnat}
\bibliography{ref}
\end{document}